\documentclass{aa}  

\usepackage{graphicx}
\usepackage{adjustbox}
\usepackage[normalem]{ulem}
\usepackage{longtable}
\usepackage{booktabs}
\usepackage{txfonts}
\usepackage{hyperref}
\defcitealias{Schnitzeler_2019}{S19}
\begin{document}

   \title{The Galactic latitude dependency of Faraday complexity in the S-PASS/ATCA RM catalogue}

\titlerunning{The Galactic latitude dependency of Faraday complexity in the S-PASS/ATCA RM catalogue}

   \author{S. Ranchod\inst{1}\thanks{E-mail: sranchod@mpifr-bonn.mpg.de}, S. A Mao\inst{1}, R. Deane\inst{2,3}, S. S. Sridhar\inst{4}, A. Damas-Segovia\inst{1}, J. D. Livingston\inst{1}, Y. K. Ma\inst{5}
          }
    \authorrunning{S. Ranchod et al.}

   \institute{Max-Planck-Institut f\"{u}r Radioastronomie, 
              Auf dem H\"{u}gel 69, 
              53121 Bonn, Germany
        \and Wits Centre for Astrophysics, School of Physics, University of the Witwatersrand, 1 Jan Smuts Avenue, Johannesburg, 2000, South Africa
        \and Department of Physics, University of Pretoria, Private Bag X20, Pretoria 0028, South Africa
        \and SKA Observatory, Jodrell Bank, Lower Withington, Macclesfield, SK11 9FT, United Kingdom 
        \and Research School of Astronomy \& Astrophysics, Australian National University, Canberra, ACT 2611, Australia
            }

   \date{Received December 18, 2023; accepted March 14, 2024}

 
  \abstract{
   The S-band Polarisation All Sky Survey (SPASS/ATCA) rotation measure (RM) catalogue is the largest broadband RM catalogue to date, increasing the RM density in the sparse southern sky. Through analysis of this catalogue, we report a latitude dependency of the Faraday complexity of polarised sources in this catalogue within 10$^\circ$ of the Galactic plane towards the inner Galaxy. In this study, we aim to investigate this trend with follow-up observations using the Australia Telescope Compact Array (ATCA). We observe 95 polarised sources from the SPASS/ATCA RM catalogue at 1.1 – 3.1 GHz with ATCA's 6 km configuration. We present Stokes QU fitting results and a comparative analysis with the SPASS/ATCA catalogue. We find an overall decrease in complexity in these sources with the higher angular resolution observations, with a complexity fraction of 42\%, establishing that the majority of the complexity in the SPASS/ATCA sample is due to the mixing-in of diffuse Galactic emission at scales $\theta > 2.8'$. Furthermore, we find a correlation between our observed small-scale complexity $\theta < 2.8'$ and the Galactic spiral arms, which we interpret to be due to Galactic turbulence or small-scale polarised emission. These results emphasise the importance of considering the maximum angular scale to which the observations are sensitive in the classification of Faraday complexity; the effect of which can be more carefully investigated with SKA-precursor and pathfinder arrays (e.g. MeerKAT and ASKAP). }

   \keywords{Polarization -- Radio continuum: galaxies -- Galaxy: general -- Techniques: polarimetric
               }

   \maketitle
%

\section{Introduction}
Magnetic fields have an essential dynamic role in galaxy evolution, particularly in influencing the distribution of cold and ionised gas and the formation of stars \citep[for a review, see][]{Beck_2015}. It is therefore important to fully understand the structure and strength of magnetic fields in galaxies. For example, there remain unanswered questions concerning the role of magneto-ionic turbulence in galaxy evolution and the correlation length of turbulence in the interstellar medium (ISM). A primary observational tool for probing magnetic fields in galaxies is the Faraday rotation of linearly polarised synchrotron emission from a background source as it passes through an intervening magneto-ionic medium \citep[e.g.][]{Lamee_2016, Ma_2020, Eyles_2020}. This background synchrotron radiation is typically from active galactic nuclei (AGN) and is parameterised by the Faraday rotation measure (RM) or, in the general case, characterised by the Faraday depth $\phi$ \citep{Burn_1966},
\begin{equation}\label{eq:phi}
    \phi(L) = 0.812 \int_{\boldsymbol{s} = 0}^{L} n_\mathrm{e} \boldsymbol{B}\cdot\mathrm{d}\boldsymbol{s} \;\; \mathrm{rad \,m}^{-2},
\end{equation}
where $n_\mathrm{e}$ and $\boldsymbol{B}$ are, respectively, the electron density and magnetic field of intervening media along the line of sight $\boldsymbol{s}$, from the source to the observer $L$. A linearly polarised synchrotron-emitting source is considered Faraday simple when the entire source is subjected to a single $\phi$ only (i.e. a single Faraday thin polarised component is detected). Meanwhile, Faraday complex sources are composed of multiple polarisation components, each experiencing a different amount of Faraday rotation due to the distribution of the magneto-ionic medium between them \citep[e.g.][]{Farnsworth_2011, Anderson_2015}. Such changes can either occur in the medium within or surrounding those sources, or in the Galactic foreground. In Faraday complex cases, the Faraday depth parameter allows for the characterisation of emitting Faraday components or magnetised sources through techniques such as RM synthesis \citep{Brentjens_2005} and Stokes QU fitting \citep{Farnsworth_2011,OSullivan_2012}. Faraday complexity is an observational classification and can be affected by various observational parameters, such as angular resolution, frequency, and telescope sensitivity \citep[e.g.][]{Sun_2015}. 

Extragalactic sources (EGSs) exhibit a variation in complex Stokes Q and U spectra as a function of $\lambda^2$, which encode the physical magnetic field properties of the emitting sources and intervening Faraday screens \citep[e.g.][]{Farnsworth_2011,OSullivan_2012}. With the advent of broadband receivers and backends in centimetre-wavelength radio telescopes -- for example, the Australia Telescope Compact Array (ATCA), Karl G. Jansky Very Large Array (VLA), Australian Square Kilometre Array Pathfinder (ASKAP), and MeerKAT -- we are, in principle, able to disentangle the various contributions in the polarised emission of Faraday complex sources by fitting multi-component physical models to broadband Q and U spectra. This reveals rich properties of the magneto-ionic medium that were previously inaccessible with narrow-band data \citep[e.g.][]{OSullivan_2017,Pasetto_2018,Ma_2019a}. Methods such as QU fitting \citep{Farnsworth_2011,OSullivan_2012} and RM synthesis \citep{Brentjens_2005} enable a good characterisation of Faraday complexity in EGSs; however, the field is still developing and there are open questions as to how the classification of Faraday complexities can be integrated into a physical model \citep[e.g.][]{Alger_2021}.

Previous studies have reported trends in Faraday complexity due to various foreground screens. \citet{Livingston_2021} found that a large percentage (95\%) of their sample of EGSs in the Galactic centre region are Faraday complex, which is attributed to small-scale ($\sim 3$ pc) turbulence in the Galactic foreground screen driven by stellar feedback. \citet{Anderson_2015} analysed the environment of 14 Faraday complex sources at an arcminute resolution and identified several possible contributors to complexity, including foreground magneto-ionic galaxy clusters and an association of complexity with ionisation fronts near neutral hydrogen structures in the ISM.

As this paper subsequently demonstrates, it is crucial to consider observations at various angular scales to fully interpret the physical scale of the Faraday complexity. In principle, interferometric observations using long baselines can filter out the smoothly varying large-scale polarised emission (e.g. from the Galactic plane) to better disentangle the contributions of small-scale Faraday components in Faraday complex sources. Such components can be associated with small-scale magneto-ionic turbulence using the RM structure function \citep[SF;][]{Haverkorn_2008,Stil_2011}. The SF can be used to reveal the outer scale of turbulent cells and how this changes with angular scale, as was done for the Galactic plane \citep{Haverkorn_2008,Livingston_2021}, the Milky Way at high Galactic latitudes \citep[e.g.][]{Mao_2010}, and Magellanic Clouds \citep[e.g.][]{Mao_2012}. This method can also improve our understanding of turbulence within nearby galaxies and the EGSs themselves \citep{Anderson_2015,Anderson_2016,OSullivan_2017}. 

RM grids allow us to understand the variation of magnetic fields on various scales and in different cosmic environments, for example, nearby galaxies \citep[e.g.][]{Gaensler_2005,Mao_2008,Livingston_2022}, galaxy clusters \citep[e.g.][]{Anderson_2021} or the Milky Way \citep{Brown_2007,Haverkorn_2008,Hutschenreuter_2022}, dependent on the density of polarised background sources. This can be constrained by more sensitive, large-area polarised intensity surveys. Modern large area polarisation surveys at lower frequencies ($\leq$ 1.1 GHz) include the LOFAR Two-metre Sky Survey \citep[LoTSS,][]{OSullivan_2023} and Spectral and Polarisation in Cutouts of Extragalactic sources from the Rapid ASKAP Continuum Survey \citep[SPICE-RACS,][]{Thomson_2023}, with 2461 and 5818 sources respectively. The \citet{Taylor_2009} catalogue derived from NRAO VLA Sky Survey is the largest RM catalogue to date, containing $\sim40~000$ sources at DEC $> -40^\circ$, observed with two narrow bands (42 MHz) at $\sim1.4$ GHz. However, due to the limitations of narrow-band spectropoliametry, this catalogue contains inaccurate RMs due to the n$\pi$ ambiguity \citep{Ma_2019a}, and increased RM uncertainties from off-axis instrumental polarisation \citep{Ma_2019b}. The SPASS/ATCA RM catalogue \citep[][hereafter \citetalias{Schnitzeler_2019}]{Schnitzeler_2019} is currently the most extensive large-area broadband catalogue at cm-wavelengths, with $\sim 4500$ sources at DEC $<0^\circ$, covering 1--3 GHz. This catalogue has contributed to improving the Galactic Faraday sky reconstruction, by increasing the RM density to 1 RM per 5 deg$^2$ in the ``southern hole" \citep{Hutschenreuter_2022}. This unique catalogue will continue to have legacy value into the SKA-era, due to its polarised source density, sky coverage and, most importantly its broad bandwidth. It is therefore important to fully understand any systematic trends observed. However, the SPASS/ATCA catalogue has several limitations, particularly its low angular resolution, short integration time and the spectra-extraction methods used. Follow-up observations can help us better characterise these limitations to optimise the synergy and scientific output of this catalogue in the pre-SKA and SKA eras.

In this paper, we present a study of the polarisation properties of a subset of 95 sources from the SPASS/ATCA catalogue with follow-up observations from the ATCA. We investigate the dependence of Faraday complexity with Galactic latitude and how this dependency changes with higher angular resolution data. Furthermore, through the QU fitting and the analysis of Faraday spectra, we aim to discern whether this dependence is caused by mixed-in diffuse Galactic emission or small-scale turbulence in the Galactic Plane. 

This paper is organised as follows; In Sect.~\ref{sec:data}, we detail the systematics we uncovered in the SPASS/ATCA catalogue and describe our observations and data calibration methods. Section~\ref{sec:analysis} describes our methods of extracting and analysing the spectro-polarimetric data. Section~\ref{sec:results} presents the results from RM synthesis, QU fitting and Faraday complexity classification. In Sect.~\ref{sec:disc}, we discuss the physical origin of Faraday complex sources, how observational constraints affect the perception of complexity and resulting caveats regarding the SPASS/ATCA catalogue. We summarise and conclude in Sect.~\ref{sec:concl}.

\section{Data}\label{sec:data}
\subsection{The SPASS/ATCA RM catalogue}\label{sec:s19}
The SPASS/ATCA RM catalogue (\citetalias{Schnitzeler_2019}) contains $\sim4500$ polarised sources at Dec $\leq 0$ deg. The survey was observed using an unconventional snapshot mode ($\sim 36$ s on source integration time) with the ATCA in the 16 cm (1.1--3.1 GHz) band. \citetalias{Schnitzeler_2019} used the hybrid ATCA configuration H168 (excluding antenna 6) with baselines between 61 -- 192 m. This configuration was chosen to maximise $uv$-coverage for equatorial sources, and corresponds to an angular scale of 2.8' $<\theta<$ 9.0',  and a reported angular resolution of $1'\times2'$. The spectra were extracted directly from the visibilities without forming spectral image cubes to reduce processing time and the required storage space. The Stokes Q and U spectra were then fit iteratively for up to 5 Faraday-thin (simple) polarised components using the \texttt{FIRESTARTER} \citep{Schnitzeler_2017} fitting algorithm. Here, sources are considered Faraday simple if they are best fit by a single polarised component and Faraday complex if they are best fit by multiple polarised components. The details of this classification and fitting methods used are elaborated on in Sect.~\ref{sec:qufit}.

\begin{figure*}
    \centering
    \includegraphics[width=0.95\textwidth]{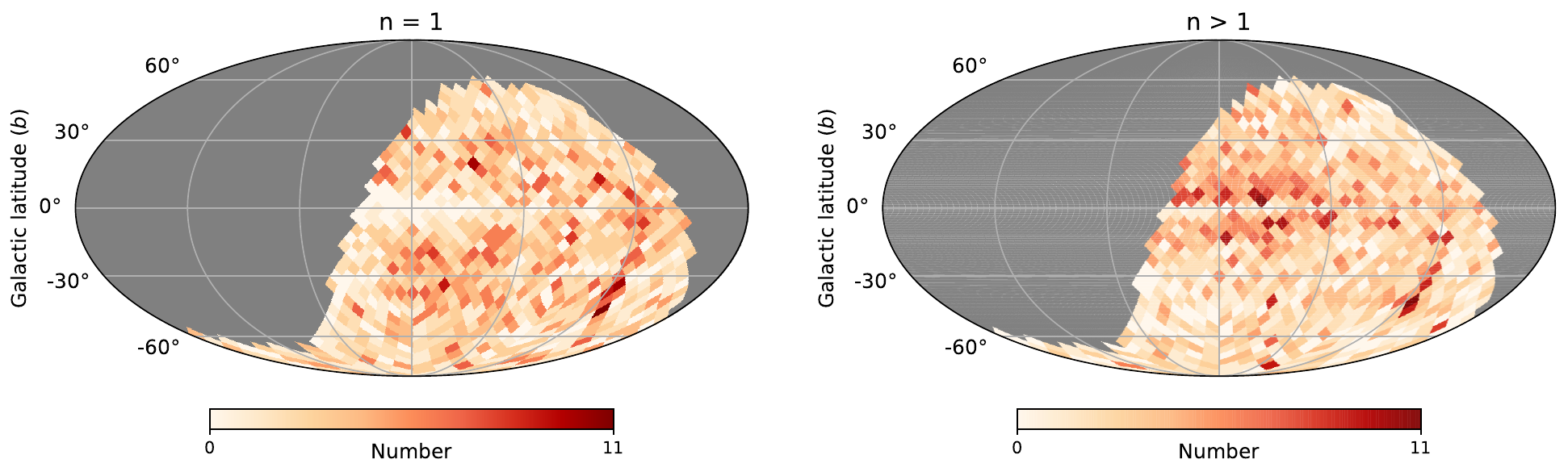}
    \caption{Sky source density distribution for Faraday simple ($n=1$, left) and Faraday complex ($n>1$, right) polarised sources from the SPASS/ATCA catalogue in Galactic Mollweide projection. The colour bar indicates the number of sources per bin.}
    \label{fig:spassatca_dist}
\end{figure*}

The spatial distribution of Faraday simple and Faraday complex source density is shown in Fig.~\ref{fig:spassatca_dist}. In this plot, we find an increase in the number of sources with multiple fitted polarised components (Faraday complex) towards the Galactic plane, as well as a decrease in the number of sources with only one fitted polarised component (Faraday simple) towards the plane. This systematic trend was not previously noted in the literature. A reduction in Faraday simple source density and a corresponding increase in Faraday complex source density is seen above and below the Galactic plane. It is more pronounced towards the inner Galaxy in Q1 and Q4 ( $30^\circ>l>0^\circ$, $360^\circ>l>270^\circ$). We also observe a deficit of Faraday simple sources at $b\sim0^\circ, l\sim280^\circ$, which is likely produced by the Gum Nebula \citep[e.g.][]{Stil_2011,Purcell_2015}. The gap in Faraday complex sources that are very close to the Galactic plane ($|b| < 1.5^\circ$), in the right panel of Fig.~\ref{fig:spassatca_dist}, is a selection effect from \citetalias{Schnitzeler_2019}.
\begin{figure}
    \centering
    \includegraphics[width=0.9\columnwidth]{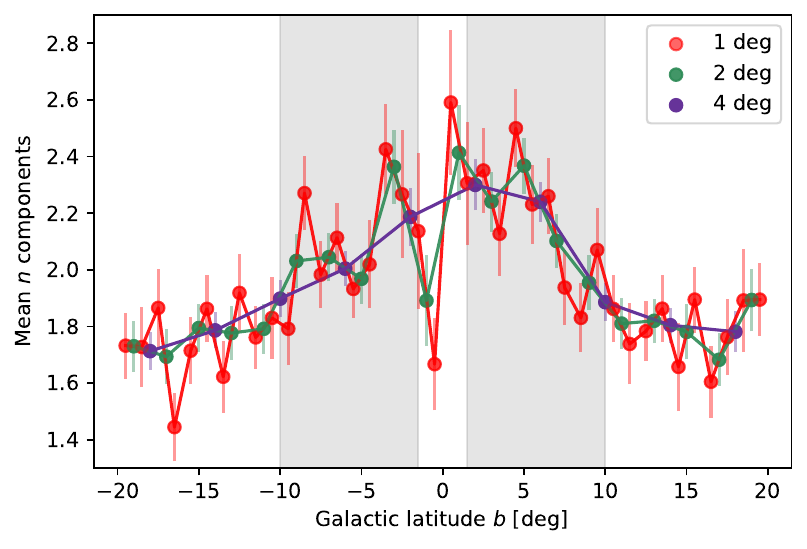}
    \caption{Mean number of fitted polarised components for sources in the SPASS/ATCA RM catalogue, as a function of Galactic latitude. The different colours indicate different latitude bins. The grey, shaded regions indicate the latitude range of our observations.}
    \label{fig:latdepend_full}
\end{figure}

We further demonstrate this trend of increased complexity in Fig.~\ref{fig:latdepend_full}, with the increase in the mean number of fitted polarised components (i.e. the Faraday complexity) as a function of Galactic latitude. The distribution is binned in latitude with bin widths of 1, 2 and 4 degrees, and a respective average of 47, 95 and 189 sources per bin. There is an apparent increase in complexity for sources closer to the Galactic plane at $|b|<10^\circ$. For larger latitude bins, this effect is more pronounced at positive Galactic latitudes. Sources that are Faraday complex, with multiple polarised components within the beam volume (i.e. along the line of sight, or within the beam resolution) that are intrinsic to the source, should be uniformly distributed across the sky without a dependency on Galactic coordinates. We propose that the increased complexity towards $b=0^\circ$ must therefore be Galactic in origin.

The short baselines and the large beam size in the \citetalias{Schnitzeler_2019} observations can be sensitive to the large-scale, diffuse polarised synchrotron emission in the Milky Way. This could be a likely scenario for added complexities in Stokes Q and U spectra close to the Galactic plane, particularly with the visibility spectra extraction method used in \citetalias{Schnitzeler_2019}. Another possible scenario for an increase in Faraday complexity could be a turbulent foreground Faraday screen caused by small-scale structures and fluctuations in electron density and magnetic fields in the Milky Way \citep[e.g.][]{Haverkorn_2008,Livingston_2021}. With higher-angular-resolution broadband observations, we aim to discern between the various scenarios to constrain the origin and scale of the complexity at low Galactic latitudes.

\subsection{ATCA observations}

We selected a sample of 105 extragalactic sources with high polarised intensities from the \citetalias{Schnitzeler_2019} catalogue for follow-up observations with the ATCA. We selected sources in close proximity to the Galactic plane, with $1.5^\circ < |b| < 10^\circ$ and $310^\circ < l < 360^\circ$, a polarised intensity lower limit of $P >$ 6 mJy and polarised fractions $p_f > 0.01$. Of the selected sources, we classify ten as extended, and 95 as unresolved by inspecting the Stokes I images from the Rapid ASKAP Continuum survey \citep[RACS;][]{McConnell_2020,Hale_2021}-low DR1, at an angular resolution of 25".

A total of 24 hours of observations were obtained over two 12-hour observing runs. The observations were done in the ATCA 16 cm band (1.1 -- 3.1 GHz) with the 6 km configuration. 48 sources ($l > 333^\circ$) were observed on 28 January 2022. Each source was observed for a total of 9.5 minutes over six (95-second) scans every $\sim 2$ hours, to optimise $uv-$coverage and parallactic angle coverage. The primary flux and bandpass calibrator PKS B1934$-$638 was observed for 20 minutes at the end of the observing run. We selected the closest secondary phase calibrators from the ATCA calibrator database. The secondary calibrators are summarised in Table~\ref{Tab:calibrators}. The six phase calibrators were observed intermittently throughout the observations, bracketing the scans of the target sources. The remaining 54 sources ($l < 333^\circ$) were observed on 5 February 2022. Each source was observed for a total of 8.2 minutes over seven (70-second) scans. We observed PKS B1934$-$638 for 10 minutes and the four phase calibrators multiple times ($\sim60$ seconds each) throughout the observation, as in the first observations. Three of the phase calibrators were also science targets (marked in Table~\ref{Tab:calibrators}).
\begin{table*}
\centering
\caption{Source names and positions of secondary phase calibrators. The calibrators also used as science targets are marked with an asterisk.}
\begin{tabular}{llcc}
\hline
Observation date & Source name & RA (J2000) & Dec (J2000) \\ \hline
28 Jan 2022      & PKS B1421$-$490    & 14:24:32   & $-$49:13:49   \\
                 & PKS B1511$-$47*    & 15:14:40   & $-$47:48:30   \\
                 & PKS B1600$-$44     & 16:04:31   & $-$44:41:31   \\
                 & PKS B1613$-$586*   & 16:17:17   & $-$58:48:07   \\
                 & PKS B1511$-$55     & 15:15:12   & $-$55:59:33   \\
                 & PKS B1352$-$63     & 13:55:46   & $-$63:26:42   \\
5 Feb 2022       & PKS B1600$-$44     & 16:04:31   & $-$44:41:32   \\
                 & PKS B1622$-$310    & 16:25:55   & $-$31:08:08   \\
                 & PKS B1759$-$39*    & 18:02:43   & $-$39:40:08   \\
                 & PKS B1740$-$517    & 17:44:25   & $-$51:44:44   \\ \hline
\end{tabular}
\label{Tab:calibrators}
\end{table*}
With our chosen array configuration, we have baselines from 192~m~--~6~km, resulting in angular scales of about $5.5"<\theta<2.8'$. With a lower maximum angular-scale sensitivity than \citetalias{Schnitzeler_2019}, and a higher angular resolution, we aim to filter out any large-scale diffuse polarised emission from the Galactic plane and minimise beam depolarisation effects. These observations have a factor $\sim$14 longer integration time than \citetalias{Schnitzeler_2019}, and we expect a factor $\sim$4 lower thermal noise. These conditions allow us to distinguish whether the \citetalias{Schnitzeler_2019} observations have been contaminated by diffuse polarised synchrotron Galactic emission and to identify localised small-scale regions of magnetic field structure, for example, turbulence, patchy polarised emission.

\subsection{Calibration and imaging}\label{sec:calim}
Data calibration was carried out following the standard procedure for Compact Array Broadband Backend (CABB) data with the \textsc{Miriad} software \citep{Sault_1995}. PKS B1934$-$638 was used for bandpass, polarisation leakage and flux calibration. The calibrators listed in Table~\ref{Tab:calibrators} were used for complex gain calibration. We did systematic manual and automatic flagging on all calibrators before and after calibration, using \textsc{Miriad} tasks \texttt{blflag} and \texttt{pgflag}. We flagged 100 MHz at the band edges due to poor sensitivity.

\begin{figure}
    \centering
    \includegraphics[width=0.8\columnwidth]{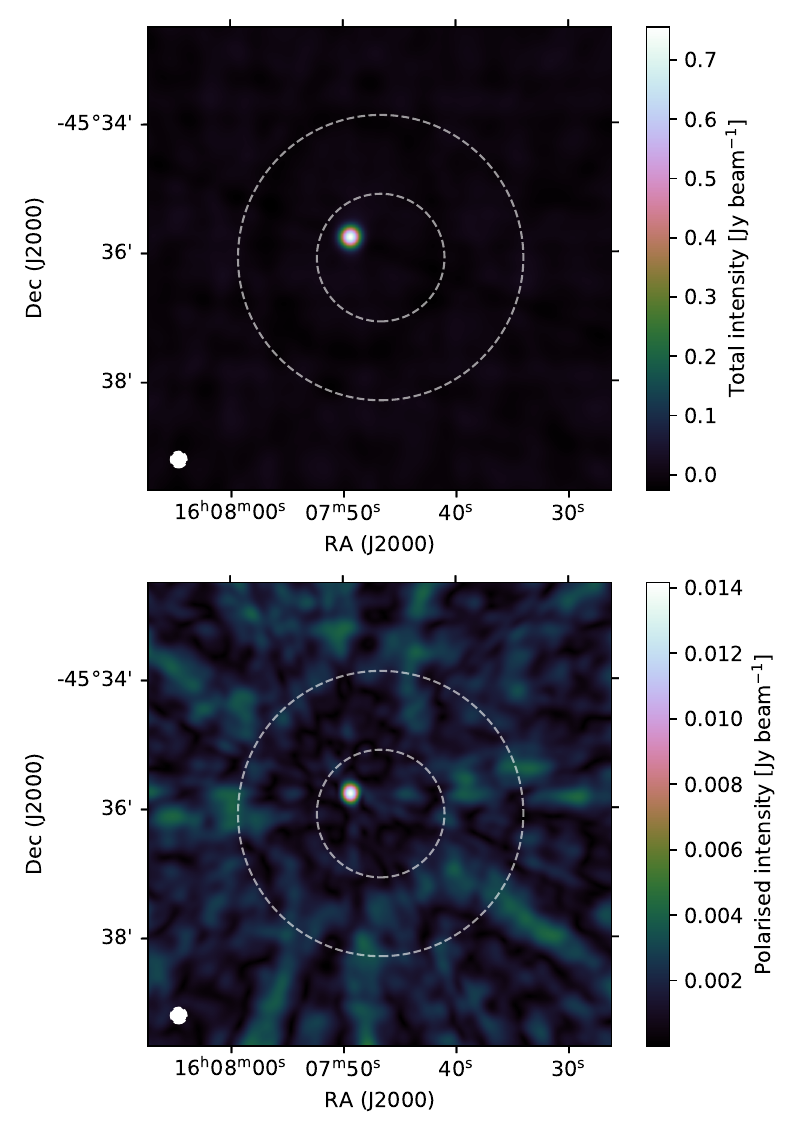}
    \caption{Example of Stokes I (top) and polarised intensity (bottom) MFS images from our observations for source SPASS~J160700$-$452802. The dashed circles show the major and minor axes of the synthesised beam for the \citet{Schnitzeler_2019} data. The restoring beam for the example images is shown in the bottom left corner (15"$\times$15")}
    \label{fig:imeg}
\end{figure}

We created shallow Stokes I Multi-Frequency Synthesis (MFS) images for all target sources using \textsc{Miriad}'s \texttt{invert}, \texttt{clean} and \texttt{restore}. Masks were generated at a 10$\sigma$ threshold, and phase and amplitude self-calibration was performed with \textsc{Miriad}'s \texttt{selfcal}. We used masked deconvolution to produce full-Stokes MFS images. The image sizes were 2000$\times$2000 pixels with a pixel size of 0.7" and we used a Briggs \citep{Briggs_1995} robust weighting of 0.5. We applied primary beam correction and smoothed to a common beam of 15"$\times$15", to account for beam variation across the 2~GHz bandwidth. This resolution corresponds to the major axis of the beam at the low-frequency end of the band, and at this angular resolution, 46 sources are spatially unresolved, and 49 have another spatial component within the 2' range of the \citetalias{Schnitzeler_2019} beam. An example image of source SPASS~J160700$-$452802 is shown in Fig.~\ref{fig:imeg}. The Stokes~V MFS images have a mean rms of $\sim 97\,\mu$Jy beam$^{-1}$, in comparison to the theoretical rms of $\sim 80\,\mu$Jy beam$^{-1}$. 

We imaged spectral cubes in Stokes I, Q, and U with \textsc{Miriad}, aiming to maximise the per-channel signal-to-noise ratio (S/N) with broader channel width but also considering the corresponding increase in bandwidth depolarisation effects. In selecting an optimal channel width, we assumed a reasonable depolarisation limit of 95\% \citep{Gaensler_2001}. Considering our lowest unflagged frequency channel at 1.3 GHz and the highest $|\mathrm{RM}|$ from the \citetalias{Schnitzeler_2019} selected sample, the maximum channel width with negligible bandwidth depolarisation is $\Delta\nu=17$ MHz. Therefore, we correspondingly formed spectral cubes with frequency steps of 17 MHz. The per-channel image sizes were 2000$\times$2000 pixels with a pixel size of 0.7" and we used a Briggs robust weighting of 0.5. As with the MFS images, we applied primary beam correction and smoothed to a common beam of 15"$\times$15".

In addition to the 95 target sources, we observed ten known extended sources, as identified from RACS cutouts, to test the scale to which the selected array configuration can filter out diffuse, extended emission. The RACS images have an angular resolution of 25'' $\times$ 25''. We find that these sources show no smoothly varying diffuse emission at scales $>3'$ in total intensity or polarisation, and that this emission has successfully been filtered out in our observations. In some cases, there is small-scale patchy emission visible in the Stokes I MFS images. This emission all originates from the lobes of resolved radio galaxies and is not Galactic. We measure the peak flux density within the area of the \citetalias{Schnitzeler_2019} synthesised beam, and find on average that only 5\% of the \citetalias{Schnitzeler_2019} total intensity flux density is detected.

\section{Spectro-polarimetric analysis}\label{sec:analysis}
\subsection{Total intensity and polarised data extraction}
\begin{figure}
    \centering
    \includegraphics[width=0.9\columnwidth]{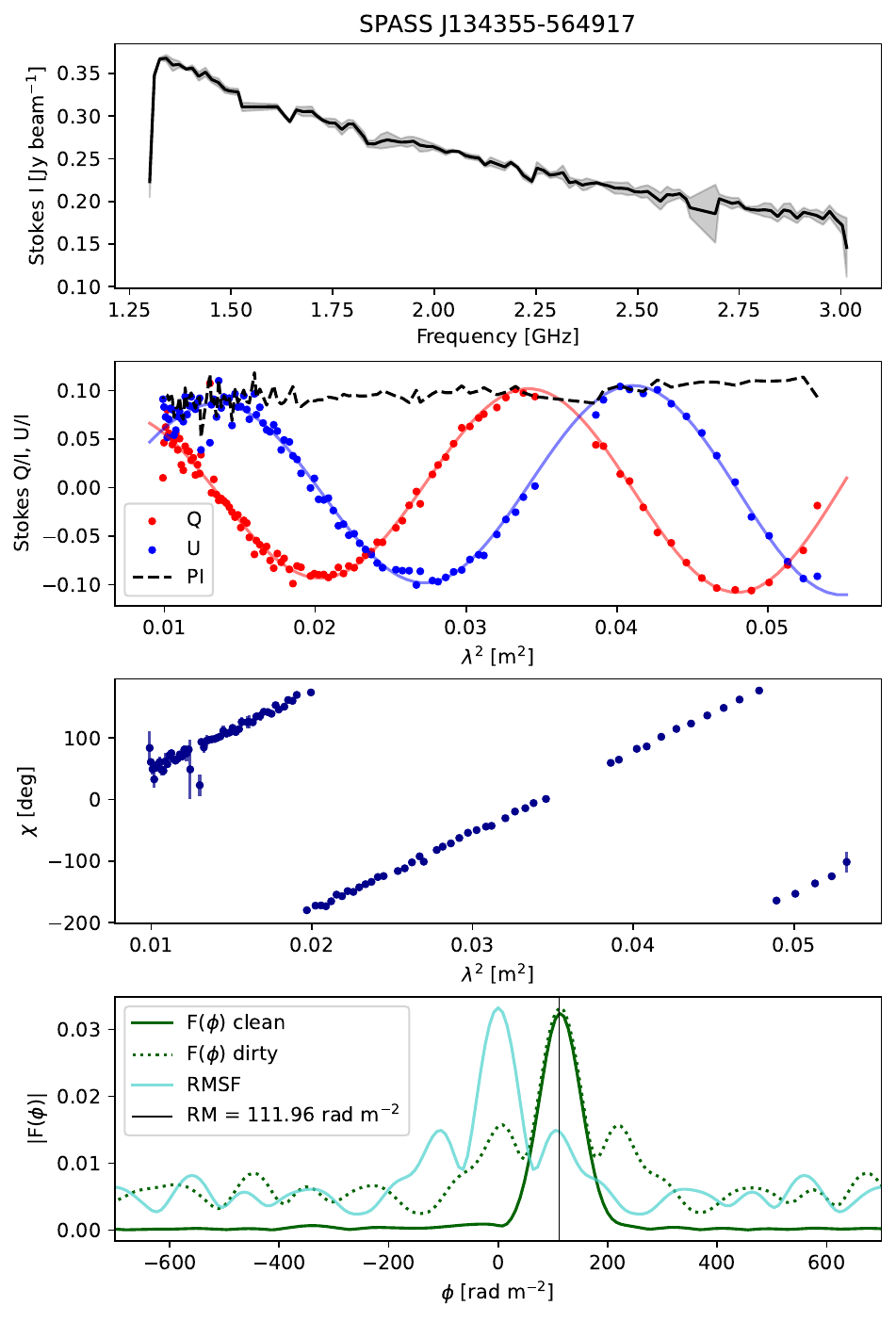}
    \caption{Example of the spectra for source SPASS~J134355$-$564917. From top to bottom: Panel 1 shows the total intensity spectrum as a function of frequency. Panel 2 shows the Stokes Q/I and U/I spectra as a function of $\lambda^2$, with the fractional polarisation spectrum indicated by the dashed line. The fitted models for Stokes Q/I and U/I are also shown. In panels 1 and 2, the shaded regions indicate the uncertainties. Panel 3 shows the polarisation angle as a function of $\lambda^2$. Panel 4 shows the dirty (green dotted) and cleaned (green solid) Faraday spectrum with the Rotation Measure Spread Function (RMSF) in cyan. The black vertical line shows the peak Faraday depth. }
    \label{fig:eg_spec}
\end{figure}

We used two different methods of extracting the total intensity flux density values. Firstly, we determined the broadband total intensity $I_\mathrm{MFS}$ by fitting 2D Gaussians to the Stokes I MFS images using \textsc{Casa} \citep{McMullin_2007} \texttt{imfit}. Similarly, we fitted 2D Gaussians to the 2100 MHz frequency slice of the Stokes I image cubes to determine the total intensity at 2100 MHz $I_{2100}$, for a more direct comparison to \citetalias{Schnitzeler_2019}. Secondly, we extracted spectra directly from the Stokes I, Q, and U spectral cubes from the position of the peak pixel in the Stokes I MFS images. We determined the RMS of the spectra from a 30" $\times$ 30" box at an angular distance of $\sim8'$ from the pointing centre. Polarised intensity $PI$ spectra were computed as 
\begin{equation}
    PI(\lambda^2) = \sqrt{Q(\lambda^2)^2 + U(\lambda^2)^2}.
\end{equation}
We disregard the correction for noise bias due to the high signal-to-noise ratios of polarised intensity in our observations (S/N $>10\sigma$, across the band). It should be noted that the $PI$ spectra are only used for visual inspection during QU fitting, and we report our polarised intensity measurements and methods in Sect.~\ref{sec:fluxdensity}. An example of the Stokes I, Q, U and polarised intensity spectra for source SPASS~J134355$-$564917 are shown in Fig~\ref{fig:eg_spec}.

\subsection{QU fitting}\label{sec:qufit}
We use QU fitting \citep{Farnsworth_2011,OSullivan_2012} to characterise the magneto-ionic environment of the emitting source as well as any polarised components within the beam volume. This is done by fitting physical models describing the local and intervening magneto-ionic environments to the polarised emission from a source (i.e. Stokes Q, U as a function of $\lambda^2$). Broadband observations offer a unique advantage for this. In particular, we want to characterise the complexity of sources in the sample based on the model with the best fit. Here, we assume all targets to be synchrotron-emitting point sources, with the possibility of the presence of additional synchrotron-emitting Faraday-thin components within the beam volume. This may include cases where the magnetised plasma is intrinsic to the point source, or the Faraday-rotating foreground screen(s) is Faraday thick (see Sect.~\ref{sec:rmsynth}) or turbulent. Before fitting, the Stokes Q and U spectra are divided by the Stokes I spectrum to remove all first-order spectral index effects. Following \citet{Schnitzeler_2017} and \citetalias{Schnitzeler_2019}, we fitted models describing the sum of up to five Faraday thin components:
\begin{align}
    q (\lambda^2) = \sum^n_{j=1}q_{j} \left(\frac{\lambda^2}{\lambda_\mathrm{ref}{^2}}\right)^{-\delta \alpha_j/2}\mathrm{cos}[2(PA_j+\phi_j \lambda^2)] \label{eq:qfit} \\
    u (\lambda^2) = \sum^n_{j=1}u_{j} \left(\frac{\lambda^2}{\lambda_\mathrm{ref}{^2}}\right)^{-\delta \alpha_j/2}\mathrm{sin}[2(PA_j+\phi_j \lambda^2)], \label{eq:ufit}
\end{align}
where $u = U/I$ and $q = Q/I$, $PA$ is the intrinsic polarisation angle, $\phi$ is the rotation measure, $\delta\alpha$ is the remaining spectral index of a given polarised component, and $\lambda_\mathrm{ref}$ is the wavelength at which the spectral index is determined. We fitted all sources with $n = [1,2,3,4,5]$. The derivation of this model and any assumptions made are detailed in Appendix~\ref{app:qufit}. With this simplified model, any intrinsic complexity within the source or turbulence in intervening Faraday screens will be modelled as separate polarised components. We use this schema in lieu of more complex models for a direct comparison of complexity classification with  \citetalias{Schnitzeler_2019}. Here, a linearly polarised synchrotron-emitting source is considered Faraday simple when there is a single polarised component within the beam volume ($n=1$), and Faraday complex when there are multiple or composite polarised media ($n>1$). We compute the Bayesian Information Criterion (BIC) to select the best-fit model. For two respective models, a difference in BIC greater than 10 represents ``very strong'' evidence in favour of the model with the smaller $n$. Given this definition, we select the model with the lowest BIC as the preferred model, except if the difference between the two lowest BIC values is less than 10. In this case, the model with the smallest $n$ is selected. An example of the fitted QU-model for source SPASS~J134355$-$564917 is shown in Fig.~\ref{fig:eg_spec}.

\subsection{RM synthesis}\label{sec:rmsynth}
The complex polarised intensity can be expressed as a function of Faraday depth $\phi$, defined in Eq.~\ref{eq:phi},
\begin{equation}
    PI(\lambda^2) = Q(\lambda^2) +iU(\lambda^2) = \int ^\infty _{-\infty} F(\phi)\exp(2i\phi\lambda^2)\mathrm{d}\phi.
\end{equation}
We note that $F(\phi)$ is the Faraday spectrum (historically referred to as the Faraday dispersion function), the complex polarised intensity at a given Faraday depth \citep{Burn_1966,Brentjens_2005}. Sources are considered Faraday simple when all polarised emission is at a single value of $\phi$ (i.e. the Faraday spectrum is a delta function). Physically, this could interpreted as a single polarised component, limited by the given $\lambda^2$ sampling. Sources are considered Faraday complex when they emit at multiple $\phi$ values. In this case, there are multiple peaks at different $\phi$ in their Faraday spectrum (i.e. multiple polarised components). From the \citetalias{Schnitzeler_2019} catalogue, we expect a large fraction of our sample to be Faraday complex. 

The observed Faraday spectrum $\tilde{F}(\phi)$ is the convolution of $F(\phi)$ with the Fourier transform of a weighted sampling function in the $\lambda^2$ domain, called the RM spread function (RMSF). We require a deconvolution algorithm to approximate the true $F(\phi)$. 
The resolution of a Faraday spectrum and the precision of $\phi$ measurements determined through RM synthesis are characterised by three quantities, $|\phi_\mathrm{max}|$ the maximum observable Faraday depth, $\delta\phi$ the resolution in Faraday space, and the maximum scale in Faraday space to which one is sensitive. These quantities are linked to the scale and resolution of the QU-spectra \citep{Brentjens_2005,Dickey_2019}. We determine the theoretical measurements of these quantities to be $|\phi_\mathrm{max}| =1732~\mathrm{rad}\,\mathrm{m}^{-2}$, $\delta\phi = 92~\mathrm{rad}\,\mathrm{m}^{-2}$ and max-scale$\,=317~\mathrm{rad}\,\mathrm{m}^{-2}$ for our observation setup. After flagging, calibration and RMsynthesis, we find the median observed FWHM of the RMSF to be $\delta\phi_\mathrm{obs} = 80~\mathrm{rad}\,\mathrm{m}^{-2}$.

We used RM synthesis \citep{Brentjens_2005} with RM clean \citep{Heald_2009} to determine the $|F(\phi)|$, RMSF and the peak Faraday depth(s) of the sources. RM synthesis was performed using \texttt{RM-Tools} \citep[Version 1.1.1,][]{Purcell_2020} from the Canadian Initiative for Radio Astronomy Data Analysis (CIRADA) tools. We used a clean threshold of $3\sigma$ for RM clean, and elect to divide the Stokes Q and U spectra by a polynomial fit to the Stokes I spectrum. Throughout this work, peak Faraday depth $\phi$ and RM are used interchangeably.


\section{Results}\label{sec:results}
Of the 105 sources, we excluded extended sources, due to the expected non-detection of large-scale emission. Our final sample consisted of 95 sources. The brightest source in the sample is SPASS~J160017$-$464922, with $I_\mathrm{MFS} = 5.1\pm0.1~\mathrm{Jy}$ and the mean Stokes I flux density is $I_\mathrm{MFS} = 0.62\pm0.08~\mathrm{Jy}$. We present a catalogue of all measured polarisation parameters as well as the results of QU fitting. The catalogue is per the RMTable2023 standards presented in \citet{Vaneck_2023}, and is available as part of the consolidated RMTable catalogue (v1.2.0) \footnote{Source catalogue also available at: \url{https://doi.org/10.17617/3.NRFFQB}.}. This includes the following quantities and their associated errors, total intensity $I_\mathrm{MFS}$, polarised intensity, $\phi_\mathrm{peak}$. The catalogue also includes $\phi_j$ as specified in Equations~\ref{eq:qfit} and \ref{eq:ufit}. An abridged version of the catalogue is shown in Appendix~\ref{app:cat}.

\subsection{Verification of SPASS/ATCA catalogue}
To verify the methods used in \citetalias{Schnitzeler_2019} and to assess our measurements, we compare our results to those from \citetalias{Schnitzeler_2019}. In this section, we compare flux density measurements and RMs.

\subsubsection{Flux density measurements}\label{sec:fluxdensity}
\begin{figure}
    \centering
    \includegraphics[width=0.9\columnwidth]{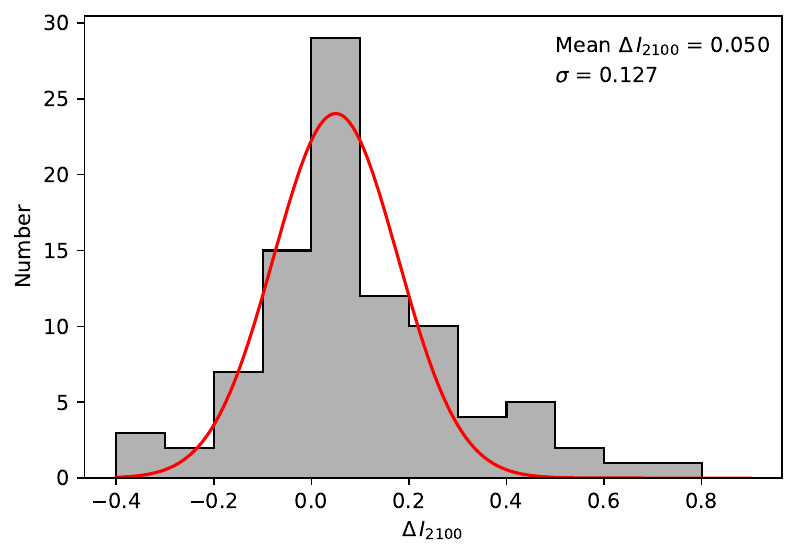}
    \caption{Histogram of $\Delta I_{2100}$, defined in Eq.~\ref{eq:I2100}. A Gaussian fit is plotted in red, with the resultant mean and standard deviation shown on the plot. The bin width is 0.1}
    \label{fig:fluxcomp_2100}
\end{figure}

\begin{figure}
    \centering
    \includegraphics[width=0.9\columnwidth]{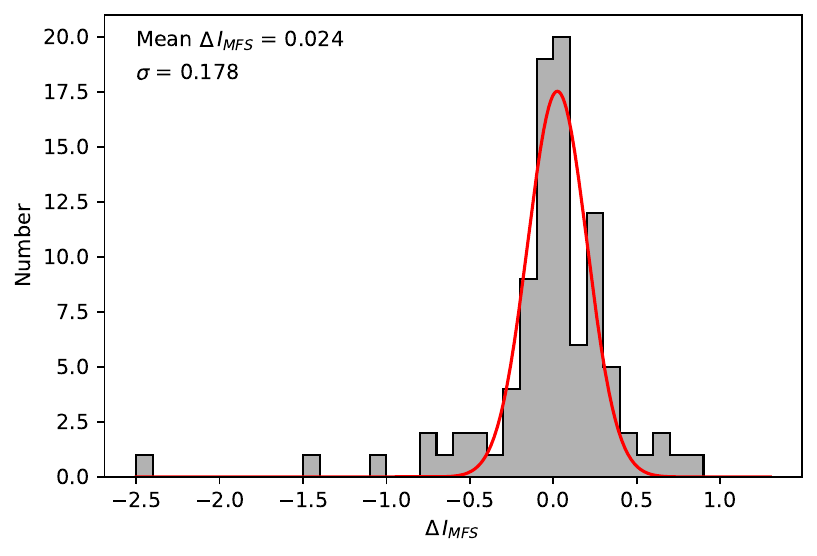}
    \caption{Histrogram of $\Delta I_\mathrm{MFS}$, defined in Eq.~\ref{eq:Imfs}. A Gaussian fit is plotted in red, with the resultant mean and standard deviation shown on the plot. The bin width is 0.1}
    \label{fig:fluxcomp_mfs}
\end{figure}

We consider the fractional difference between the Stokes I flux density measurements in our work and various catalogues in the literature. Fig.~\ref{fig:fluxcomp_2100} shows a fractional comparison $\Delta I_{2100}$ between the total intensity flux density from the \citetalias{Schnitzeler_2019} catalogues $I_\mathrm{S19}$ and the measured flux density at 2100 MHz, where
\begin{equation}\label{eq:I2100}
    \Delta I_{2100} = \frac{I_{2100} - I_\mathrm{S19}}{I_{2100}}.
\end{equation}
We find these flux density measurements broadly consistent, with a mean $\Delta I_{2100} = 0.050 \pm 0.018$ (i.e. 5\% of $I_{2100}$). We do expect a non-zero residual due to the difference in $uv$-coverage between observations. We also compare the measured MFS total intensity flux densities with the RACS catalogue \citep{Hale_2021}, assuming a spectral index of $\alpha = -0.7$ to interpolate the RACS flux densities from 887 MHz to 2100 MHz, $I_\mathrm{RACS,2100}$. Similar to Eq.~\ref{eq:I2100}, we define the fractional comparison as
\begin{equation}\label{eq:Imfs}
    \Delta I_\mathrm{MFS} = \frac{I_\mathrm{MFS} - I_\mathrm{RACS,2100}}{I_\mathrm{MFS}}.
\end{equation}
A histogram of $\Delta I_\mathrm{MFS}$ is shown in Fig.~\ref{fig:fluxcomp_mfs}. We find the flux density measurements to be consistent, with a mean $\Delta I_\mathrm{MFS} = 0.024 \pm 0.018$ (i.e. 2\% of $I_\mathrm{MFS}$). In doing a fractional comparison of the measured flux densities of $I_\mathrm{MFS}$ and $I_\mathrm{2100}$, we find an inconsistency, with an 11\% mean fractional residual. Further investigation into the effect of broadband imaging on flux density measurements and how this compares to visibility spectra is beyond the scope of this work, given our objectives.

For the purpose of the catalogue, we determined the $PI$ of a source as the sum of the peak amplitudes in the Faraday spectrum. We find a broad agreement between the measured $PI$ and the reported values in \citetalias{Schnitzeler_2019}, with a mean fractional residual of $0.06\pm0.03$. The spread of the fractional residuals is significantly larger than for the Stokes I flux densities (Fig.~\ref{fig:fluxcomp_2100}), with $\sigma = 0.30$. Furthermore, 18 sources have fractional residuals beyond $3\sigma$. The majority of the outliers have low polarised intensity at $PI<20$ mJy. Here, we see a deviation from the consistency with \citetalias{Schnitzeler_2019}, with larger $PI$ measurements in \citetalias{Schnitzeler_2019}. On account of different observational setups, we expect these inconsistencies in polarised intensity due to the filtering of and sensitivity to different physical scales of linearly polarised emission. 

\subsubsection{Rotation measures}
\begin{figure}
    \centering
    \includegraphics[width=\columnwidth]{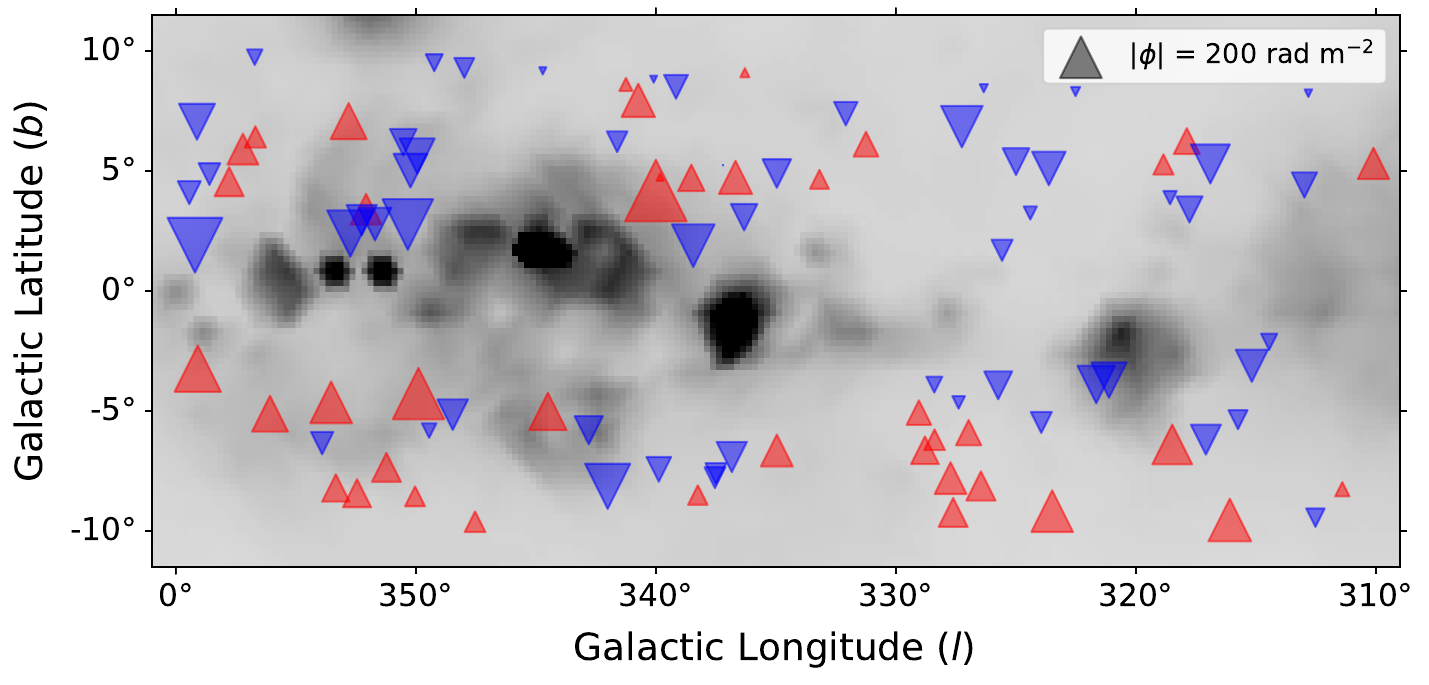}
    \caption{Spatial distribution of observed sources plotted on the Wisconsin H-Alpha Mapper (WHAM) Sky Survey \citep{Haffner_2003} H$\alpha$ map of the Galactic plane (greyscale). The marker size is proportional to the amplitude of the measured peak Faraday depth of the source, as indicated in the legend. Red upward triangles are positive $\phi$ and blue downward triangles are negative $\phi$.}
    \label{fig:RM_dist}
\end{figure}

\begin{figure}
    \centering
    \includegraphics[width=0.9\columnwidth]{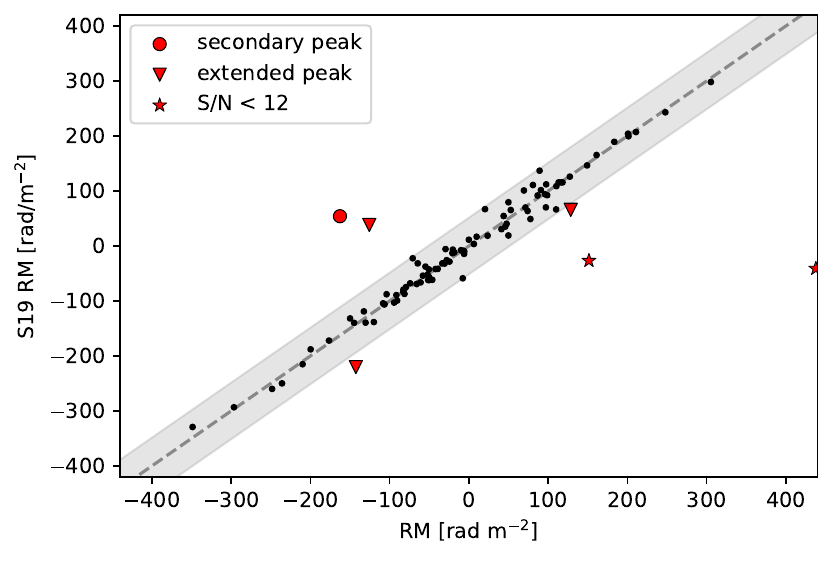}
    \caption{Comparison between the measured $\phi$ from this work and those from \citetalias{Schnitzeler_2019}. The uncertainties on our $\phi$ measurements are typically of order $\sim 1$ rad m$^{-2}$ and are not visible on the scale of this plot. The shaded region indicates 5$\sigma$ of the residuals. The outliers beyond this are plotted in accordance with the explanation in the main text. Cases (i), (ii) and (iii) are plotted as red stars, circles and triangles, respectively.}
    \label{fig:rm_compare}
\end{figure}

The mean $\phi$ of the sample is -$4.50\pm13~\mathrm{rad}\,\mathrm{m}^{-2}$, the mean $|\phi|$ is $99\pm13~\mathrm{rad}\,\mathrm{m}^{-2}$ and the maximum $|\phi|$ is 437.73~$\mathrm{rad}\,\mathrm{m}^{-2}$. The spatial distribution of our sources and their peak Faraday depths are plotted in Fig.~\ref{fig:RM_dist}. Fig.~\ref{fig:rm_compare} shows a comparison between our measurements and those from the \citetalias{Schnitzeler_2019} catalogue. We find the majority of $\phi$ measurements consistent with the \citetalias{Schnitzeler_2019} measurements, with six outliers beyond 5$\sigma$. We examine the Faraday spectra of the 5$\sigma$ outliers on a case-by-case basis and identify three causes of this inconsistency. These causes are listed below and are indicated in Fig.~\ref{fig:rm_compare}. 
\begin{enumerate}[(i)]
    \item sources with low S/N peak Faraday depth (i.e. depolarised in our observations). Here, sources have S/N $< 12$, that is, 10\% of the mean S/N of the sample.
    \item sources where our peak Faraday depth measurement corresponds to a secondary peak in the \citetalias{Schnitzeler_2019} Faraday spectra.
    \item sources which are Faraday thick in the \citetalias{Schnitzeler_2019} Faraday spectra, or are composed of multiple unresolved peaks. In this case, our determined $\phi$ from RM synthesis lies within the Faraday depth range of the \citetalias{Schnitzeler_2019} extended structure but has an offset from the maxima.
\end{enumerate}   


\subsection{Galactic latitude dependence of Faraday complexity}\label{ssec:compclass}
\begin{figure}
    \centering
    \includegraphics[width=0.9\columnwidth] {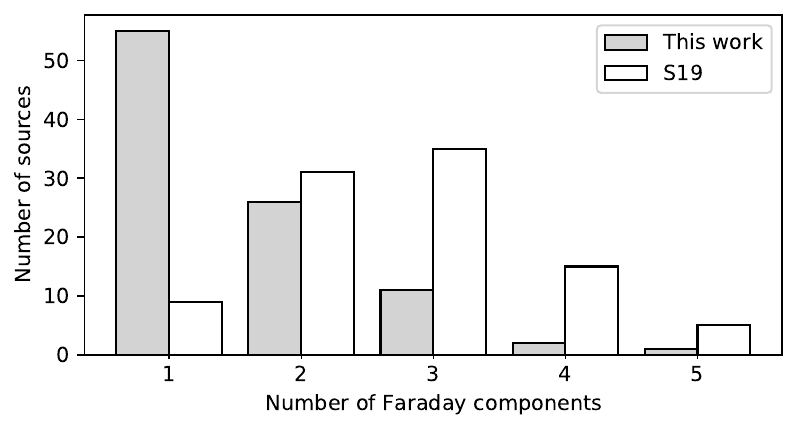}
    \caption{Histogram comparing the number of sources best fit by a given QU fitting model, as described in Eqs.~\ref{eq:qfit} and \ref{eq:ufit}, for this work (grey) and \citetalias{Schnitzeler_2019} (white).}
    \label{fig:hist_n_comp}
\end{figure}

In this section, we present the Faraday complexity classification of sources using the QU fitting method outlined in Sect.~\ref{sec:qufit}. We find that 40 (42\%) sources are complex, which is less than half of the 90\% complex sources found for the same sample in \citetalias{Schnitzeler_2019}. The number of sources for a given number of Faraday components in comparison to \citetalias{Schnitzeler_2019} is shown in Fig.~\ref{fig:hist_n_comp}. 

\begin{figure}
    \centering
    \includegraphics[width=0.96\columnwidth]{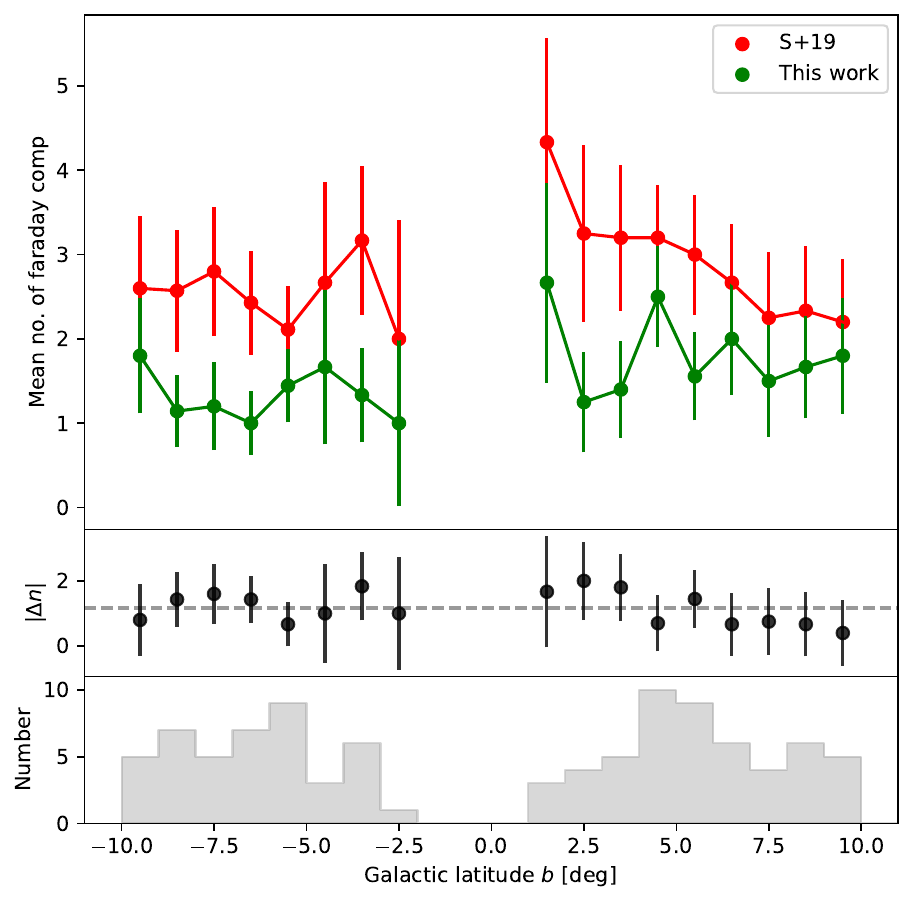}
    \caption{The average number of fitted Faraday components of the observed sources (green) and the same sources from \citetalias{Schnitzeler_2019} (red) as a function of Galactic latitude. The errorbars are Poisson uncertainties.  The central panel shows the difference in the average number of components per bin between this work and \citetalias{Schnitzeler_2019}. The mean difference ($|\Delta n| = 1.1\pm0.1$) is plotted as the dashed line. The bottom panel shows the number of sources in each 1-degree bin.}
    \label{fig:lat_depend}
\end{figure}

In Fig.~\ref{fig:lat_depend}, we plot the average number of Faraday components as a function of Galactic latitude. It should be noted that our sample is unevenly split on either side of the Galactic plane, with 52 sources above the plane and 43 sources below the plane. We find an overall decrease in the number of components per latitude bin in this work compared to \citetalias{Schnitzeler_2019}, with a mean difference in number of components as $|\Delta n| = 1.1\pm0.1$. We find the slope of the increase of the average number of components towards the Galactic plane to be lower below the plane in both samples, which is also observed for the full S-PASS/ATCA catalogue, as shown in Fig.~\ref{fig:latdepend_full}. In our measurements, we also find that the average number of components is higher at positive latitudes than at $b<0^\circ$. We note in Sect.~\ref{sec:calim} that 49 sources have a secondary spatial component within the \citetalias{Schnitzeler_2019} beam. For these sources, the mean $|\Delta n| = 1.32\pm0.1$, and for the unresolved sources, we find $|\Delta n| = 0.96\pm0.1$. The increased $|\Delta n|$ for sources with multiple components suggest that the exclusion of these spatial components from our spectra does contribute to our observed reduction of complexity with respect to \citetalias{Schnitzeler_2019}. However, the $|\Delta n|\sim1$ for unresolved sources indicates that this is not the dominant effect for the reduced number of fitted components in Fig.~\ref{fig:lat_depend}.

We return to the scenarios introduced in Sect.~\ref{sec:s19} for the cause of increased Faraday complexity towards the Galactic plane, first considering the case for small-scale turbulence. It is unlikely that small-scale turbulence has a strong latitude-dependent effect on complexity at the angular resolution of the \citetalias{Schnitzeler_2019} survey. 
Because we have an overall consistency of $\phi$ with the \citetalias{Schnitzeler_2019} catalogue, we do not spatially resolve turbulence within the Galactic plane foreground. In the cases of multiple spatial components in the \citetalias{Schnitzeler_2019} beam, we also find consistent $\phi$ with both the \citetalias{Schnitzeler_2019} catalogue and the brightest component, with a median difference in $\phi$ of 7 rad$\,$m$^{-2}$, well within 5$\sigma$ in Fig.~\ref{fig:rm_compare}. \citet{Basu_2019} show through MHD simulations that Faraday depth varies smoothly along the line of sight due to spatially correlated structures in the magnetic field, and that the scale of magnetic field variation is consistent with the driving scale of turbulence. If the increase of complexity in \citetalias{Schnitzeler_2019} is purely due to Galactic plane turbulence, we would observe the same number of Faraday components between this work and \citetalias{Schnitzeler_2019}. 

In the case for mixed-in Galactic diffuse emission, with the low angular resolution \citetalias{Schnitzeler_2019} observations ($\sim2'$), approximately 64 times more diffuse flux density would be detected than in our observational setup. This would result in \citetalias{Schnitzeler_2019} detecting one additional Faraday component from diffuse emission, as observed (Fig.~\ref{fig:lat_depend}). Because the average number of Faraday components decreases by one in our observations, it is likely that the filtered-out components correspond to large-scale regions of polarised emission from the Galactic plane at scales $> 2.8'$, corresponding to 2.4 pc at a 3 kpc distance.  This follows the assumption that the polarisation angle of the extended Galactic foreground does not vary on large scales, and that polarised intensity remains approximately constant. We can therefore conclude that mixed-in polarised Galactic synchrotron emission is the dominant cause for the increase of the number of Faraday components towards the Galactic plane observed in \citetalias{Schnitzeler_2019}, a trend that we no longer observe in observations with a smaller maximum angular scale. Despite the overall decrease in complexity, there remains complexity in 42\% of the sources. We suggest that these complexities are due to turbulence in the Galactic plane or polarised emission at scales $< 2.8'$, and examine these sources on a case-by-case basis in Sect.~\ref{sec:disc}. 

It should be noted that Fig.~\ref{fig:lat_depend} shows an increase in complexity at $b\sim5^\circ$. This is caused by a bias in observed sources, where we have an overdensity of sources that are within 5 deg of longitude of the Galactic centre at $b\sim5^\circ$ (see Fig.~\ref{fig:complexdist}). If we consider Fig.~\ref{fig:lat_depend} excluding sources with $l>355^\circ$, we no longer observe this trend. 

\section{Discussion}\label{sec:disc}
 In the following section, we outline the various physical mechanisms that introduce Faraday complexity within a beam volume, and discuss how this can be interpreted or perceived differently depending on the observational setup and telescope limitations. In Sect.~\ref{sec:originofcomp} we examine the origin of complexity in our observed sample in detail.
\subsection{The perception of complexity}
\begin{figure}
    \centering
    \includegraphics[width=0.94\columnwidth]{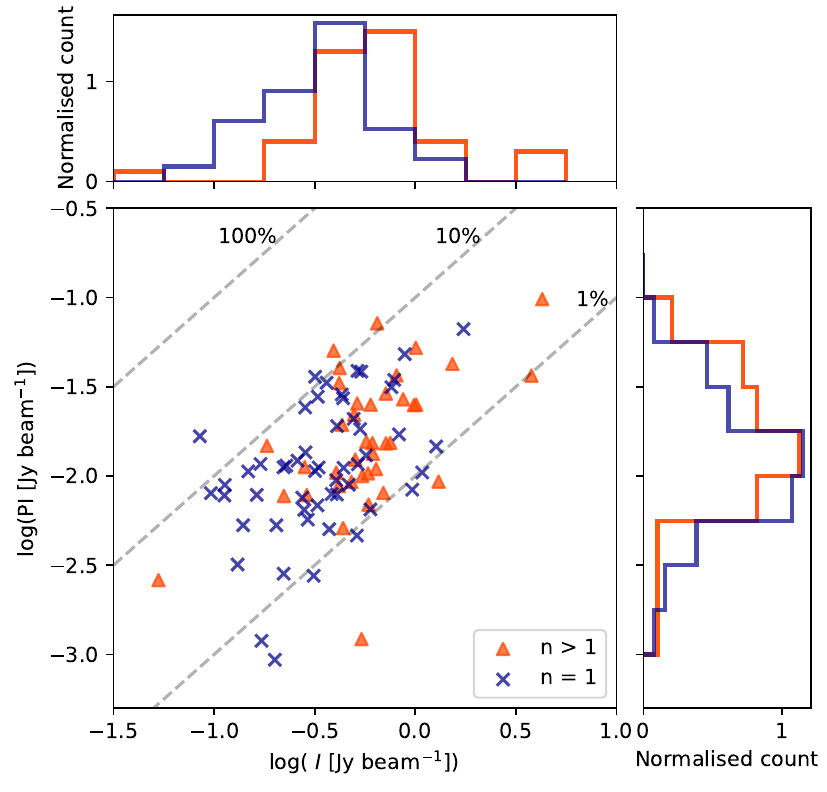}
    \caption{Scatter plot of the polarised intensity and the MFS total intensity flux densities. Faraday simple sources are plotted as blue crosses and Faraday complex sources are plotted as orange triangles. The grey dashed lines represent constant fractional polarisation, labelled as a percentage. The histograms above (to the right of) the plot shows the distribution of the total intensity (polarised intensity flux densities) for the Faraday simple and complex sources, in blue and orange, respectively.}
    \label{fig:fracpol}
\end{figure}
In Fig.~\ref{fig:fracpol} we plot the total intensity vs. polarised intensity, along with histograms for the distributions for Faraday simple and complex sources. The majority of sources have a fractional polarisation between 1\% and 10\%. We find that Faraday complex sources have preferentially higher flux density values in both polarised and total intensity. This shows that it is more likely for a source to be classified as complex if it has a higher S/N polarised intensity (based on the chosen method of classification). For fainter sources in polarised intensity, possible secondary components may not be detected with sufficient S/N, and these sources would preferentially be classified as simple. This bimodality is more evident in total intensity and is consistent with \citet{Anderson_2015}, who found through simulations that there is a limit to the detection of complex sources with S/N. Although there are significantly lower S/N levels in the \citetalias{Schnitzeler_2019} catalogue, we observe an overall decrease in the number of complex sources due to the lower resolution and the sensitivity to larger maximum angular scales in \citetalias{Schnitzeler_2019}.

\begin{figure}
    \centering
    \includegraphics[width=0.8\columnwidth]{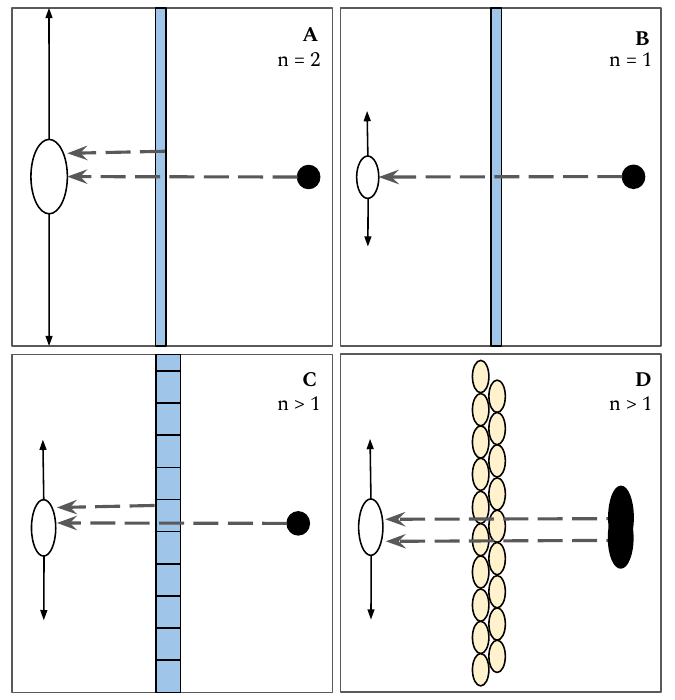}
    \caption{Diagram illustrating the effect of multi-resolution observations on the interpretation of Faraday complexity. The black circle indicates a linearly polarised, synchrotron-emitting source, spatially unresolved in A, B and C. The white ellipse indicates the angular resolution of the observations, and the vertical arrows indicate the maximum angular scale of the observations. The blue-shaded regions are the polarised emitting foreground,  with emission at smaller scales in panel C, and the yellow-shaded region is a turbulent foreground screen. The dashed arrows show the paths of emission from the various sources, and the number of observed Faraday components is indicated in the top right corner.}
    \label{fig:scale-diagram}
\end{figure}

In Sect.~\ref{ssec:compclass}, we find a significant difference in complexity fractions between our observations and \citetalias{Schnitzeler_2019}. The top row in Fig.~\ref{fig:scale-diagram} (panels A and B) shows a depiction of the effect of multi-resolution observations on our perception of Faraday complexity. Both depict a linearly polarised EGS with a smooth polarised emitting foreground. Panel A describes a scenario where the observational setup has a lower angular resolution and a larger maximum angular scale (e.g. \citetalias{Schnitzeler_2019}). In this case, the telescope is sensitive to the linearly polarised emission from the observed background EGS, as well as the foreground. The polarised emission of the background source undergoes Faraday rotation when passing through the extended foreground. From the observer's point of view, there are two polarised components within the beam volume, $n = 2$ (Faraday complex), following the classification scheme in Sect.~\ref{sec:qufit}. Panel B shows an identical system with a different observational setup. Here, we have higher angular resolution, and the maximum angular scale of the observation is smaller than the extended foreground polarised emission, therefore, the majority of the foreground emission is filtered out (e.g. this work). The emission from the EGS will undergo Faraday rotation from the foreground, but only a single polarised component is detected within the beam volume, $n = 1$ (Faraday simple) within this observation setup.

For completeness, we illustrate two cases of foreground-induced complexity detectable with our observational setup. Panel C is a similar case to panel B, with a smooth, extended foreground component that has been filtered out. However, here we consider small-scale (i.e. smaller than the beam), patchy polarised emission in the foreground. The presence of such emission within the beam volume will increase the number of Faraday components, that is, $n > 1$. Panel D illustrates an example of a turbulent foreground screen. Depending on the scale of the turbulent cells \citep[e.g.][]{Livingston_2021}, various regions of polarised emission from the background source may experience different amounts of Faraday rotation, while passing through different turbulent cells. This would result in multiple polarised components being detected within the beam volume ($n > 1$, Faraday complex). To ascertain this and constrain the scale of turbulence, the extent of the background source would have to be larger than the scale of turbulence. Observations of well-modelled extended sources or a dense RM grid are required to investigate this further.

Our results emphasise the importance of observing Faraday complexity at different angular scales to disentangle various polarised components within the beam volume. The removal of the smooth large-scale emission from the foreground is crucial for an accurate interpretation of the linear polarisation signal from background sources for Galactic magnetism studies. The disregard of this can mislead our understanding of small-scale magnetic structures through observation of background EGSs. In general, the effective removal or filtering of Galactic contamination at all scales is important for determining accurate polarisation information from EGSs. The effect of maximum angular scale on Faraday complexity should be carefully considered depending on the science goals. This effect can be further investigated with broadband observations of the Galactic plane regions with more complex interferometer arrays, such as the MPIfR-MeerKAT Galactic Plane Survey, \citep[MMGPS;][]{Padmanabh_2023}. Within the footprint of this survey, it is expected that $\sim2.5\times10^4$ linearly polarised sources will be detected. Furthermore, a larger sample will enable a more robust understanding of the effect of Faraday complexity with angular scale. In comparison to cm-wavelengths, m-wavelength observations of extragalactic sources are largely affected by depolarisation, and are therefore not sensitive to Faraday complex sources. However, modern m-wavelength surveys \citep[LoTSS;][]{OSullivan_2023} can contribute to the analysis of Faraday complexity by providing a confirmed Faraday simple sample for a holistic investigation of this classification scheme.

\subsection{The origin of complexity}\label{sec:originofcomp}
In Sect.~\ref{ssec:compclass} we identified that the Faraday complexities in the observed sample must originate from polarised emission at scales $< 2.8'$ (i.e. the maximum angular scale of our observations), but it is difficult to distinguish whether the additional polarised components are intrinsic to the extragalactic source, or small-scale, patchy Galactic polarised emission \citep[e.g.][]{Schnitzeler_2009} or turbulence.
\begin{figure}
    \centering
    \includegraphics[width=\columnwidth]{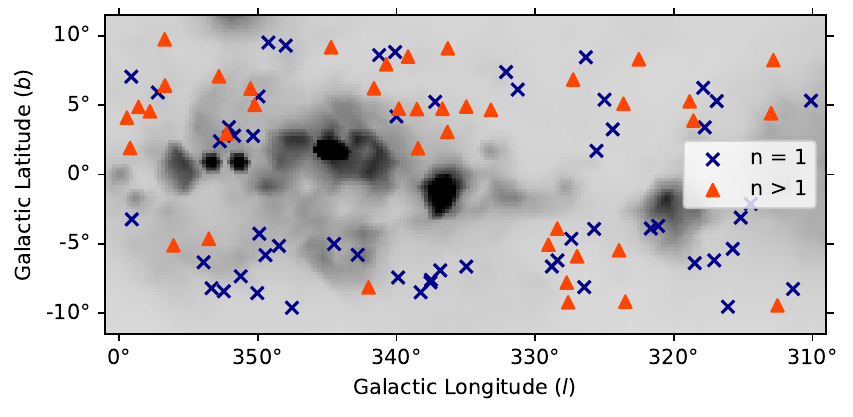}
    \caption{Spatial distribution of sources, as shown in Fig.~\ref{fig:RM_dist}. The Faraday simple sources are indicated by blue crosses and the Faraday complex sources are plotted as orange triangles.}
    \label{fig:n_compdist}
\end{figure}
Fig.~\ref{fig:n_compdist} shows the spatial distribution of Faraday simple and complex sources. From this figure, we show that the complex sources are not uniformly distributed on the sky and it is, therefore, unlikely that the observed complexity is solely intrinsic to the sources. We subsequently evaluate this quantitatively in Sect.~\ref{sec:spiral}. 

\subsubsection{Faraday spectrum classification}\label{sec:classification}
\begin{figure}
    \centering
    \includegraphics[width=0.9\columnwidth]{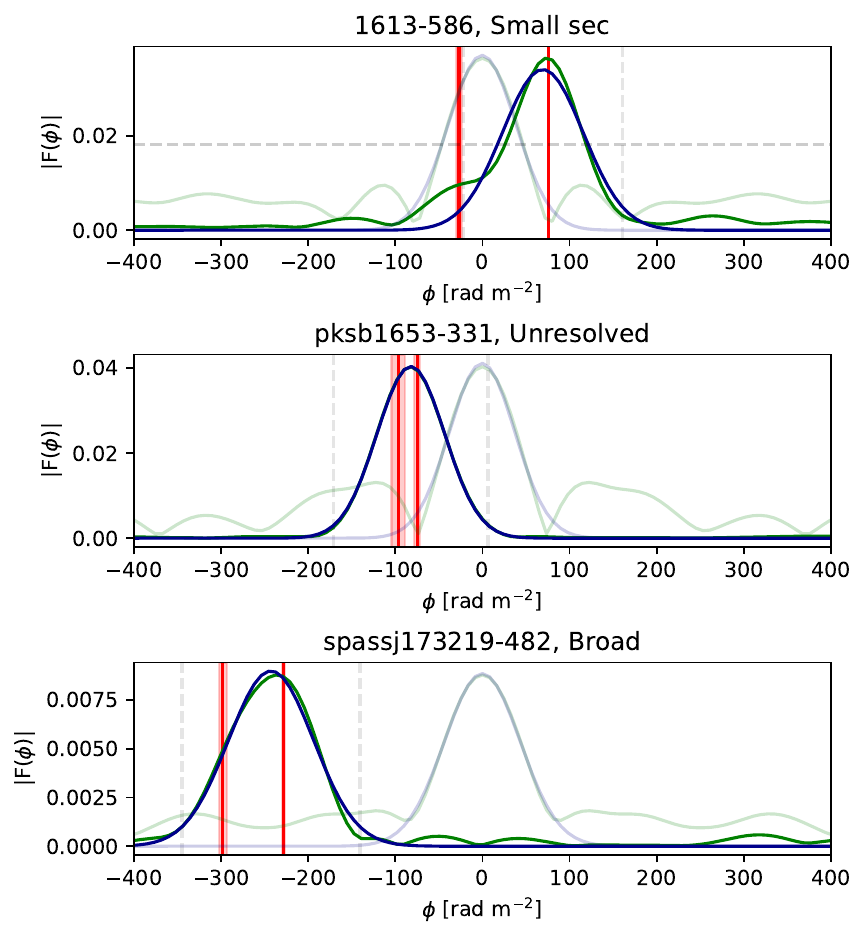}
    \caption{Examples of the Faraday spectrum classes `small sec' (top), `unresolved' (centre), and `broad' (bottom) as described in Sect.~\ref{sec:classification}. The plots show the Faraday spectrum (green) and the fitted Gaussian (blue). The same is shown for the RMSF and its Gaussian fit, in lower opacity. The red vertical lines show the fitted $\phi$ from QU fitting and their uncertainties. The 3$\sigma$ extent and the half-maximum amplitude of the fitted Gaussians are indicated with dashed grey lines.}
    \label{fig:class-eg}
\end{figure}

With the limitations of our selected QU fitting models, we further investigate the Faraday complex sources and the approximations made when fitting these models. To better understand the origin of complexity for our observed sources, we manually inspect and classify the Faraday spectra of the complex sources, also considering the limitations in the Faraday depth domain and Faraday resolution due to the observational parameters. We use four different labels to classify the complex sources: 
\begin{enumerate}[(i)]
    \item `Broad': We fitted Gaussian functions to the dominant peak in the Faraday spectrum and the RMSF. A source is considered broad if the FWHM of the Faraday spectrum peak is 10\% greater than that of the RMSF \citep[see e.g.][]{Ma_2019a}. Additionally, more than one of the $\phi$ identified in QU fitting should be within the Faraday spectrum peak. This scenario would describe an external Faraday thick source, for instance, a Faraday rotating `slab' of thermal electrons and magnetic fields overlapping with the synchrotron-emitting volume, leading to a differential Faraday rotation along the line of sight. This scenario could also describe a secondary, low amplitude peak at a $\phi <3\sigma$ of the dominant peak $\phi$, that cannot be fully resolved as a separate peak within our Faraday depth resolution. 
    \item `Unresolved': The FWHM of the dominant peak is equal to, or up to 10\% greater than the FWHM of the RMSF, and there are multiple QU fitting identified $\phi$ values within the FWHM of the peak. This classification has the same physical scenario as the broad classification, with the exception that the difference in Faraday depth between multiple peaks or the Faraday thickness of the intervening medium is unresolved in Faraday depth space with our frequency sampling.
    \item `Big sec': The secondary resolved peak, or any of the non-dominant Faraday spectrum peaks, has an amplitude $\geq$50\% of the amplitude of the dominant peak. This scenario could describe a secondary Faraday rotating and emitting source. From its high amplitude, we could assume this is likely not due to Galactic polarised emission.
    \item `Small sec': Any of the non-dominant peaks have an amplitude $<50$\% of the dominant peak.
\end{enumerate}
In this scheme, spectra can be classified with any number of these labels (e.g. broad and small sec) corresponding to properties of multiple peaks, and an example of spectra for the various classifications is shown in Fig.~\ref{fig:class-eg}. For the broadened and unresolved sources, we cannot reliably derive the amplitude of the multiple polarised components. In these cases, we consider an upper limit as the amplitude of the Faraday spectrum at the Faraday depth of the fitted RM. For sources classified as unresolved, we determine the difference in Faraday depth between the two unresolved components $|\mathrm{RM}_1 - \mathrm{RM}_2|$, for RMs obtained from QU fitting. We find 15 unresolved sources, with a mean $|\mathrm{RM}_1 - \mathrm{RM}_2|$ of $34.35\,\mathrm{rad}\,\mathrm{m}^{-2}$. We provide a further discussion on this classification in Sect.~\ref{sec:spiral}.

\subsubsection{Multiple spatial components}\label{sec:spatcomp}
In Sect.~\ref{sec:calim}, we smoothed the MFS images and image cubes to a common resolution of 15" $\times$ 15" to account for beam variation across the band. To distinguish which sources have multiple spatial components within the beam, we visually inspect the high-frequency end of the unsmoothed image cubes at 3014 MHz. These images have a typical resolution of 3.5" $\times$ 5.2". We manually identify sources with multiple components in total intensity within the smoothed beam and find 23 (i.e. 24\%) sources that have multiple spatial components. We only consider the Stokes I images here, as we are unable to reliably manually identify peaks in polarisation due to low S/N in the channel maps. A breakdown of the complexity classification of these sources is given in Table~\ref{tab:multispat}. 

We do not detect multiple bright Faraday peaks (Big sec class) in any of these sources and find that 8 (34\%) of the multi-component sources have a faint secondary Faraday peak (Small sec class). However, we cannot ascertain that the multiple Faraday components correspond to the observed multiple spatial components. Of the 23, 10 sources are Faraday simple. This is consistent with \citet{OSullivan_2017} and \citet{Ma_2019a}, who find that spatially extended sources can be Faraday simple, at angular resolutions of 1'' and 45'', respectively. An explanation for this trend could be the Laing-Garrington effect \citep{Laing_1988,Garrington_1988}, where we expect a large asymmetry in polarised intensity between the two Faraday components due to depolarisation. This may be an extreme case of the Laing-Garrington effect where one spatial component is completely depolarised. A more likely explanation is the effect of AGN lobe orientation, where both components encounter similar Faraday screens along the line of sight \citep[e.g.][]{OSullivan_2017}, and therefore have similar $\phi$. Given we have only 23 sources with multiple spatial components, we do not have a large enough sample for a deeper investigation into this trend. We would require higher resolution observations across a broad frequency range to analyse these components separately to better understand the individual systems.

We briefly discuss the case of Faraday complexity in sources with a single spatial component (i.e. spatially unresolved). The complexity classification for these sources is detailed in Table~\ref{tab:multispat}. We find that 38\% of spatially unresolved sources are Faraday complex. Although it has been shown in the literature that Faraday complexities can be linked to source morphology at a $\sim10^{-2}$~arcsec scale \citep[e.g.][]{Ma_2019a}, we would require higher resolution observations to explore this further and a statistically significant sample size to investigate the prevalence of this. Because we find a larger percentage of complex sources in spatially resolved sources (56\%) compared to spatially unresolved sources (38\%), it is certainly possible that the complexity is purely intrinsic to the EGSs. However, due to the Galactic dependency of complexity in \citetalias{Schnitzeler_2019} (Sect.~\ref{sec:s19}), it is important to investigate whether the remaining complexity in the observed sample is Galactic in origin. We do so in the following section.
\begin{table}[]
    \centering
    \caption{Number (percentage) of sources with each complexity classification and Faraday spectrum classification (Sect.~\ref{sec:classification}) for the sources with multiple spatial components, as well as the number of sources with each complexity classification for sources with a single spatial component. It should be noted that Faraday complex sources may be classified into multiple categories.}
    \begin{tabular}{lc}
    \hline
     Source classification & Number \\\hline 
     \textbf{Multiple spatial components:} & \textbf{23} \\
     \hspace{3mm}Faraday simple & 10 (44\%) \\
     \hspace{3mm}Faraday complex & 13 (56\%)\\
     \hspace{6mm} {Broad} &  2\\
     \hspace{6mm} {Unresolved} & 6\\
     \hspace{6mm} {Big sec} & 0\\
     \hspace{6mm} {Small sec} & 8\\
     \hline
     \textbf{Single spatial component:} & \textbf{72} \\
     \hspace{3mm}Faraday simple & 45 (62\%)\\
     \hspace{3mm}Faraday complex & 27 (38\%)\\
     \hline
    \end{tabular}
    \label{tab:multispat}
\end{table}

\subsubsection{Galactic ISM and complexity in the spiral arms}\label{sec:spiral}
We consider various interstellar media as possible tracers for turbulence or small-scale fluctuations in the Galactic plane. In the presence of a turbulent screen, we expect to observe depolarisation effects and detect multiple fitted polarised components, given the QU fitting models we have used. Additionally, \citet{Anderson_2015} suggest that HI and H$\alpha$ regions act as a proxy for identifying regions where complex Faraday screens are present. To investigate this, we measure the HI column densities from the HI4PI survey maps \citep{McClure-Griffiths_2009,Collab_2016}, as well as the H$\alpha$ flux densities from the WHAM H$\alpha$ map \citep{Haffner_2003}. We do not find a direct relationship between complexity and these various gas density measurements. The typical angular resolution is 14.5' for the HI4PI maps in the southern sky and 1$^\circ$ for WHAM H$\alpha$, all significantly larger than the scale at which we measure the complexity. It is therefore difficult to reliably investigate this relation without higher resolution observations. 
\begin{figure}
    \centering
    \includegraphics[width=0.95\columnwidth]{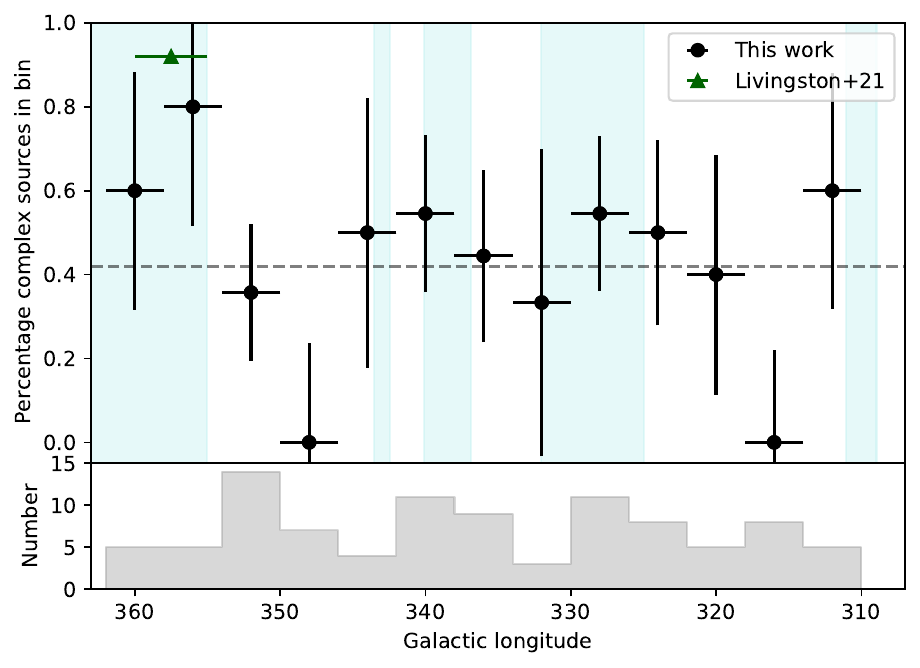}
    \caption{Percentage of complex sources for each longitude bin of 4 deg (black circles). The x error bars indicate the extent of the bin, and the y errors are determined through bootstrapping. The dashed line shows the percentage of complex sources for the full sample at 42\%. The cyan-shaded regions indicate longitudes which correspond to the Galactic spiral arms listed in the text and the longitude range within 5 degrees of the Galactic centre. The bottom panel is a histogram showing the number of sources in each longitude bin. The green triangle shows the percentage of complex sources at $355^\circ < l < 360^\circ$ in \citet{Livingston_2021}.}
    \label{fig:percentcomp}
\end{figure}
\begin{figure*}
    \centering
    \includegraphics[width=0.85\textwidth]{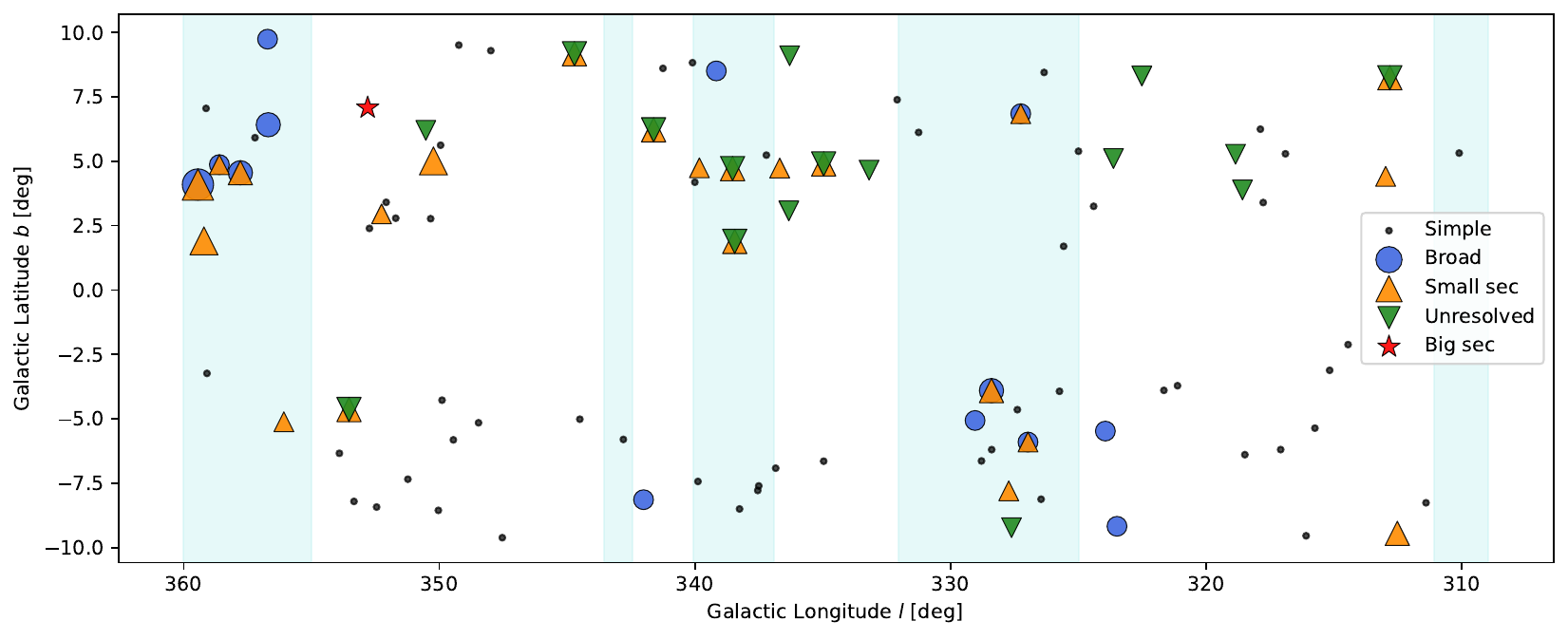}
    \caption{Spatial distribution of sources, colourised by their Faraday complexity classification, as described in Sect.~\ref{sec:classification}. The cyan-shaded regions show longitudes that correspond to the Galactic spiral arms listed in the text and the longitude range within 5 degrees of the Galactic centre. The Faraday simple sources are plotted as black dots. The size of the markers for Faraday complex sources is scaled according to the number of QU-fitted components.}
    \label{fig:complexdist}
\end{figure*}

We have investigated the Galactic latitude dependence of Faraday complexity and now consider an increase in complexity at particular Galactic longitudes between $310^\circ < l < 360^\circ$.  Our observations include the Crux-Centaurus ($l \sim 310^\circ$), Norma ($l \sim 328^\circ$), Inner Perseus ($l \sim 338^\circ$) and Inner Sagittarius ($l \sim 343^\circ$) spiral arm tangents \citep[e.g.][]{Vallee_2022} as well as the Galactic centre, where the spiral arm tangents are estimated from the broad $^{12}$CO(1-0) tracer. In Fig.~\ref{fig:percentcomp} we plot the fraction of complex sources per longitude bin. It should be noted that the longitude range for spiral arms on this figure is indicative, and we cannot constrain the full extent of the individual spiral arms. We find that in longitude bins including the spiral arms, we have a greater mean complexity than over the whole sample. Considering a comparison of complexity fractions with the literature, we also find this increased complexity within 5 degrees of the Galactic centre, as consistent with \citet{Livingston_2021} (95\% complexity). This increase could be due to the greater number of Faraday components emitting at small scales in the spiral arms at Galactic latitudes $1.5^\circ < |b| < 10^\circ$.  The observed sources in \citet{Livingston_2021} and \citet{OSullivan_2017} are suitable for a direct comparison, as they also use the 16~cm band in the ATCA with the 6~km baseline configuration, ensuring similar spatial and frequency resolution. \citet{OSullivan_2017} select an extragalactic sample of sources away from the Galactic plane, but with sources up to $|b|>20^\circ$. They find a 55\% complexity fraction. This is congruent with what we find, on average. \citet{Anderson_2015} find a complexity fraction of 12\% for sources at $b\sim55^\circ$, considering that they have a lower resolution of $\sim1'$. This complexity fraction for a non-Galactic region is consistent with the average complexity fraction we find in the interarm regions (Fig.~\ref{fig:percentcomp}).  

We determined the correlation between our binned sample and the positions of the spiral arms using the Pearson Correlation Coefficient $\rho$. Here, each spiral arm is simply modelled as a Gaussian, with the centroid and standard deviation as the centre and width of the cyan bands in Fig.~\ref{fig:percentcomp}. We included an additional component at the Galactic centre, with $x_0 = 0^\circ$ and $\sigma = 5^\circ$, and assume a complexity fraction of 100\% at the centroid of all spiral arms and the Galactic centre. We tested this by randomly assigning a boolean complexity classification to 10 000 samples with the same longitude distribution as our observed sources and a total complexity fraction of 42\%. We calculate $\rho$ for the random samples and the p-value, determined as the probability of determining a correlation at least as strong as our observed sample across the 10 000 random samples. We repeat this experiment for longitude bin widths of $\Delta l = [3, 4, 5]$ degrees. The smallest possible bin size with at least 1 source per bin is $\Delta l = 3$ deg. The results are displayed in Table~\ref{tab:correlation}. We find the strongest correlation at $\Delta l = 5$ deg with $\rho = 0.65$, and that the correlation coefficients for all bins are statistically significant with at least 97.9\% confidence. This suggests a fair correlation between increased complexity percentage and the longitudes of the Galactic spiral arms. This result is consistent with the RM structure functions in \citet{Haverkorn_2008}, who find a smaller driving scale of turbulence ($\sim1$ pc) in the spiral arm regions than in the interarm regions ($\sim 100$ pc), and suggests that the increase of complexity is attributed to turbulence at smaller scales in the spiral arms. However, a larger observed sample and a more detailed Galactic plane model are required to draw a more robust conclusion. 
\begin{table}
    \centering
    \caption{Pearson Correlation coefficient $\rho$ for the correlation between the increased complexity classification and the positions of the Galactic spiral arms for different longitude bins. The associated p-value for 10 000 randomly assigned samples is also shown.}
    \begin{tabular}{lcc}
    \hline
    $\Delta l$ bin & $\rho$ & p \\
    \hline 
    3$^\circ$   & 0.57  &  0.010 \\
    4$^\circ$   & 0.63  &  0.011 \\
    5$^\circ$   & 0.65  &  0.021 \\
    \hline
    \end{tabular}
    \label{tab:correlation}
\end{table}

To further investigate the origin of this complexity, we plot the distribution of sources, colourised by their complexity classification in relation to the spiral arms and Galactic centre (Fig.~\ref{fig:complexdist}). Through visual inspection, we find that sources classified as small sec and/or broad are concentrated around the Galactic centre or the spiral arm regions, as expected in the presence of a low flux density polarised emission. Here, it is possible that the broadening is due to smaller amplitude secondary peaks caused by small-scale emission within the Galactic plane that is unresolved at this Faraday depth resolution, or a broadening effect from small-scale Faraday depth fluctuations caused by turbulent cells. Both of these observed effects support the argument for a smaller turbulent cell size in the spiral arm regions that in the interarm regions. We discuss the cases for Faraday rotation and polarised emission separately:

\textit{Pure rotation:} In the interarm regions, we may be resolving out turbulence, that is, a low number of turbulent cells per beam volume probed. This would simply induce Faraday rotation, without increasing complexity. In the spiral arm regions, with a smaller turbulent cell size, we have an increased number of turbulent cells within the ISM volume probed by the beam. This would result in Faraday dispersion, and cause a broadening in the Faraday spectrum (Sect.~\ref{sec:classification}) and a complex classification.

\textit{Pure emission:} We would expect smoother polarised emission in the interarm regions, due to large-scale turbulence. This larger-scale emission is not detected at the scales of our observations. In the spiral arm regions, we are sensitive to the smaller scale, patchy emission, caused by the small-scale turbulence.

As discussed above, both cases or a combination of the two could contribute to the observed complexity in the spiral arm regions. However, we would require higher resolution observations in both angular resolution and Faraday depth to asses this fully.

In Fig.~\ref{fig:complexdist} we find that the majority of sources that are classified as unresolved in Faraday depth are distributed on the northern half of the plane. This is also what predominantly contributes to the increase of complexity at northern latitudes, as shown in Fig.~\ref{fig:lat_depend}. This north-south Galactic asymmetry is expected in polarised synchrotron emission \citep[e.g.][]{Vidal_2015,Robitaille_2017}. Although we find no relation between HI column density and Faraday complexity at the positions of the individual sources, we find that there is an increased average HI column density above the plane ($5.3\times10^{21}\,\mathrm{cm}^{-2}$) than below the plane ($3.1\times10^{21}\,\mathrm{cm}^{-2}$). Small-scale magneto-ionic structures (e.g. filaments at scales of $\sim 10$ pc) with low amplitude $n_\mathrm{e}\mathbf{B}$ within this extended HI region could be a possible explanation for this increased complexity \cite[e.g.][]{Clark_2019,Ma_2023}. With higher Faraday depth resolution observations and QU fitting with more advanced models, we will be able to discern whether this is an effect of unresolved Faraday dispersion by turbulence in the northern Galactic plane Faraday screen or the effect of polarised structures in Galactic HI.

\section{Conclusion}\label{sec:concl}
In this paper, we investigate the increase of Faraday complexity towards $b = 0^\circ$ in the S-PASS/ATCA RM catalogue \citep[S19;][]{Schnitzeler_2019}. We obtain follow-up observations of 95 sources at higher angular resolution, using the 6 km-baseline configuration of ATCA. We determine the peak Faraday depth $\phi$ of the sources using RM synthesis and find these measurements to be consistent with the \citetalias{Schnitzeler_2019} catalogue. By fitting Stokes Q and U spectra with models describing the sum of multiple Faraday simple components, we classify sources as Faraday simple or Faraday complex. With the higher resolution observations, we find that 42\% of the sources are complex ($n>1$), in comparison to the 90\% complexity in the same sample from \citetalias{Schnitzeler_2019}. Furthermore, we found that there is no longer a strong trend of increased Faraday complexity towards the Galactic plane, limiting the scale of complexity to $\theta < 2.8'$ (2.4 pc at a 3 kpc distance), the maximum angular scale of our observations. From this, we constrain that the increase in complexity in the \citetalias{Schnitzeler_2019} catalogue can be attributed to, on average, an additional polarised components from mixed-in diffuse Galactic emission at scales $\theta > 2.8'$. 

To investigate the remaining complexity at smaller scales, we further classify Faraday complex sources based on the characteristics in their Faraday spectra. We find tentative evidence that there is an above-average complex source percentage at longitudes corresponding to the Galactic spiral arms, suggesting turbulence or small-scale patchy emission within these regions. In addition, we observe a trend in an overdensity of sources with $|\mathrm{RM}_1 - \mathrm{RM}_2|$ smaller than the Faraday spectrum resolution in the northern Galactic plane. This could be attributed to low amplitude $n_\mathrm{e}\mathbf{B}$ fluctuations in the HI disk of the Milky Way. 

Overall, we find the total intensity and $\phi$ measurements in the \citetalias{Schnitzeler_2019} catalogue reliable and consistent with our observations, with the caveat that particular care should be taken regarding the angular resolution and the maximum angular scale when comparing polarised intensity measurements, QU fitting results, and Faraday complexity between various observations at $|b|<10^\circ$.

\begin{acknowledgements}
We thank Rainer Beck for helpful discussions surrounding this work and feedback on the manuscript, as well as Aritra Basu for QU fitting scripts. SR acknowledges the support from the International Max Planck Research School (IMPRS) for Astronomy and Astrophysics at the Universities of Bonn and Cologne. RPD acknowledges funding from the South African Research Chairs Initiative of the Department of Science and Innovation and National Research Foundation (Grant ID 77948). The Australia Telescope Compact Array is part of the Australia Telescope National Facility (https://ror.org/05qajvd42) which is funded by the Australian Government for operation as a National Facility managed by CSIRO. We acknowledge the Gomeroi people as the Traditional Owners of the Observatory site. The data used in this work will be shared upon reasonable request to the authors. 
\end{acknowledgements}

%
%

\bibliographystyle{aa}
\bibliography{example}

\appendix

\section{QU fitting models}\label{app:qufit}
In this appendix, we detail the key assumptions made in the derivation of the QU fitting model. As stated in Sect.~\ref{sec:qufit}, we consider a model describing the sum of up to five Faraday thin components, where for a given component, the polarised intensity is
\begin{align}
    P(\lambda^2) = P_0 \exp \left(2i(\phi\lambda^2 + PA)\right),
\end{align}
where $P_0$ is the polarised intensity, $\phi$ is the Faraday depth, and $PA$ is the polarisation angle. Because we are fitting the polarisation fraction $p(\lambda^2)$, we divide by by the total intensity spectrum,
\begin{align}
    p(\lambda^2) = \frac{P(\lambda^2)}{I(\lambda^2)}, 
\end{align}
assuming $I(\nu)$ to be a power law as a function of frequency $\nu$:
\begin{align}
    I(\nu) = I_0 \left(\frac{\nu}{\nu_0}\right)^{-\alpha},
\end{align}
where $I_0$ is the total intensity flux density at a reference frequency $\nu_0$ and $\alpha$ is the spectral index. This can be written as,
\begin{align}
    I(\lambda^2) = I_0 \left(\frac{\lambda^2}{\lambda_0^2}\right)^{\alpha/2},
\end{align}
resulting in
\begin{align}\label{eq:fitapp}
    p(\lambda^2) = p_0\left(\frac{\lambda^2}{\lambda_0^2}\right)^{-\alpha/2} \exp \left(2i(\phi\lambda^2 + PA)\right).
\end{align}
As specified in the text, dividing the Q and U spectra by the Stokes I spectrum before fitting, will account for the dominant spectral index, (i.e. the spectral index of the background source $\alpha_0$). This model serves to fit any variation in spectral index as contributed from emitting components within the beam volume, for which Equation~\ref{eq:fitapp} can be rewritten as
\begin{align}
    p(\lambda^2) = p_0\left(\frac{\lambda^2}{\lambda_0^2}\right)^{-\delta\alpha/2} \exp \left(2i(\phi\lambda^2 + PA)\right),
\end{align}
where $\alpha = \alpha_0 +\delta\alpha$.

\section{Source catalogue}
\label{app:cat}

\begin{table*}
    \centering
    \caption{Catalogue of the 95 analysed sources. Column (1) is the source ID from the \citetalias{Schnitzeler_2019} catalogue. Columns (2)--(5) have the positional information of each source. Columns (6)--(7) contain the total intensity measurements and columns (8)--(11) contain polarisation properties of each source. Sources identified to have multiple spatial components within the 15''$\times$15'' beam (see Sect.~\ref{sec:spatcomp}) are marked with an asterisk.}
    \begin{adjustbox}{width=\textwidth,center}
    \begin{tabular}{llllllllllcc}
    \hline
    & \citetalias{Schnitzeler_2019} ID (1) & RA (2) & Dec (3) & $l$ (4) & $b$ (5) & $I_\mathrm{MFS}$ (6) & $I_{2100}$ (7) & PI (8)& $\phi_\mathrm{peak}$ (9) & $n$ (10) & $n$ (11)\\
    & & (J2000) & (J2000) & (deg) & (deg) & (mJy) & (mJy) & (mJy) & rad m$^{-2}$ & (this work) & (\citetalias{Schnitzeler_2019}) \\
    \hline
1 & NVSSJ161021-395858 & 16h10m17s & $-39$d58m58.0s & 339.162 & $8.507$ & 599$\pm$4 & 613$\pm$7 & 25.08$\pm$0.13 & $-65.82\pm$0.16 & 2 & 2 \\
2 & NVSSJ161234-390625 & 16h12m36s & $-39$d06m59.0s & 340.093 & $8.827$ & 221$\pm$6 & 208$\pm$14 & 11.17$\pm$0.10 & $-5.85\pm$0.35 & 1 & 1 \\
3 & NVSSJ161737-382827* & 16h17m36s & $-38$d28m32.0s & 341.249 & $8.606$ & 172$\pm$6 & 163$\pm$9 & 1.19$\pm$0.10 & $+20.40\pm$3.17 & 1 & 3 \\
4 & SPASSJ143426-623949 & 14h34m26s & $-62$d39m49.8s & 314.448 & $-2.118$ & 431$\pm$5 & 403$\pm$6 & 28.63$\pm$0.12 & $-30.62\pm$0.15 & 1 & 2 \\
5 & NVSSJ171148-333841 & 17h11m48s & $-33$d38m42.0s & 352.078 & $3.407$ & 200$\pm$2 & 225$\pm$9 & 0.93$\pm$0.11 & $+110.09\pm$4.68 & 1 & 4 \\
6 & SPASSJ144346-631714* & 14h43m46s & $-63$d17m14.0s & 315.168 & $-3.116$ & 568$\pm$15 & 495$\pm$26 & 13.05$\pm$0.22 & $-119.86\pm$0.38 & 1 & 1 \\
7 & NVSSJ170918-352521* & 17h09m18s & $-35$d25m24.0s & 350.337 & $2.77$ & 1736$\pm$24 & 1687$\pm$30 & 66.55$\pm$0.60 & $-296.36\pm$0.11 & 1 & 3 \\
8 & SPASSJ144453-552959* & 14h44m53s & $-55$d29m59.5s & 318.578 & $3.887$ & 503$\pm$21 & 400$\pm$32 & 12.35$\pm$0.15 & $-20.78\pm$0.35 & 2 & 3 \\
9 & NVSSJ175445-401157 & 17h54m43s & $-40$d12m16.0s & 351.229 & $-7.344$ & 326$\pm$38 & 209$\pm$43 & 6.85$\pm$0.12 & $+96.26\pm$0.56 & 1 & 2 \\
10 & NVSSJ162732-353731* & 16h27m31s & $-35$d37m37.0s & 344.712 & $9.172$ & 544$\pm$10 & 479$\pm$9 & 9.98$\pm$0.17 & $-6.04\pm$0.48 & 3 & 3 \\
11 & NVSSJ171257-280935* & 17h12m55s & $-28$d09m29.0s & 356.689 & $6.418$ & 4275$\pm$94 & 3766$\pm$77 & 97.83$\pm$0.80 & $+52.87\pm$0.09 & 3 & 3 \\
12 & SPASSJ153448-604059 & 15h34m48s & $-60$d40m60.0s & 321.652 & $-3.894$ & 228$\pm$11 & 170$\pm$14 & 11.43$\pm$0.19 & $-162.65\pm$0.39 & 1 & 5 \\
13 & NVSSJ163736-330905 & 16h37m34s & $-33$d09m9.0s & 347.982 & $9.294$ & 274$\pm$3 & 267$\pm$6 & 7.56$\pm$0.09 & $-46.16\pm$0.50 & 1 & 2 \\
14 & NVSSJ171309-341830 & 17h13m09s & $-34$d18m39.0s & 351.703 & $2.789$ & 598$\pm$4 & 616$\pm$9 & 6.50$\pm$0.11 & $-125.65\pm$0.71 & 1 & 3 \\
15 & NVSSJ171652-254925 & 17h16m50s & $-25$d48m58.0s & 359.123 & $7.052$ & 312$\pm$23 & 290$\pm$37 & 2.76$\pm$0.14 & $-149.86\pm$1.39 & 1 & 2 \\
16 & NVSSJ171405-334539* & 17h14m04s & $-33$d45m51.0s & 352.258 & $2.955$ & 281$\pm$4 & 252$\pm$12 & 11.23$\pm$0.21 & $-106.89\pm$0.55 & 2 & 5 \\
17 & SPASSJ155724-603116 & 15h57m24s & $-60$d31m16.3s & 323.943 & $-5.476$ & 288$\pm$21 & 251$\pm$19 & 7.82$\pm$0.11 & $-51.38\pm$0.56 & 2 & 2 \\
18 & NVSSJ171609-280135 & 17h16m08s & $-28$d01m37.0s & 357.208 & $5.915$ & 203$\pm$23 & 361$\pm$95 & 5.29$\pm$0.15 & $+110.46\pm$0.84 & 1 & 5 \\
19 & SPASSJ160009-581104 & 16h00m09s & $-58$d11m4.7s & 325.737 & $-3.931$ & 760$\pm$21 & 685$\pm$21 & 31.45$\pm$0.24 & $-90.51\pm$0.16 & 1 & 3 \\
20 & NVSSJ171736-334208 & 17h17m35s & $-33$d42m17.0s & 352.731 & $2.394$ & 786$\pm$15 & 869$\pm$17 & 34.55$\pm$0.18 & $-248.25\pm$0.15 & 1 & 2 \\
21 & NVSSJ173725-394608* & 17h37m24s & $-39$d46m18.0s & 349.892 & $-4.275$ & 404$\pm$7 & 369$\pm$13 & 9.48$\pm$0.11 & $+305.37\pm$0.45 & 1 & 1 \\
22 & NVSSJ174840-365150 & 17h48m38s & $-36$d51m55.0s & 353.531 & $-4.639$ & 751$\pm$39 & 770$\pm$123 & 15.33$\pm$0.15 & $+201.40\pm$0.31 & 3 & 3 \\
23 & SPASSJ175715-414936 & 17h57m15s & $-41$d49m36.0s & 350.035 & $-8.554$ & 965$\pm$31 & 774$\pm$72 & 8.37$\pm$0.15 & $+45.92\pm$0.69 & 1 & 4 \\
24 & SPASSJ160017-464922* & 16h00m17s & $-46$d49m22.5s & 333.185 & $4.655$ & 5064$\pm$104 & 4616$\pm$98 & 429.88$\pm$1.70 & $+41.12\pm$0.04 & 2 & 3 \\
25 & NVSSJ175622-312215 & 17h56m20s & $-31$d22m26.0s & 359.087 & $-3.236$ & 437$\pm$11 & 396$\pm$13 & 27.28$\pm$0.30 & $+247.91\pm$0.21 & 1 & 3 \\
26 & NVSSJ175657-372326 & 17h56m57s & $-37$d23m42.0s & 353.907 & $-6.338$ & 162$\pm$7 & 141$\pm$13 & 7.83$\pm$0.15 & $-57.96\pm$0.54 & 1 & 2 \\
27 & NVSSJ175659-345422* & 17h56m58s & $-34$d54m37.0s & 356.079 & $-5.113$ & 615$\pm$17 & 565$\pm$14 & 13.31$\pm$0.21 & $+149.15\pm$0.38 & 2 & 2 \\
28 & NVSSJ180347-384729 & 18h03m51s & $-38$d47m39.0s & 353.336 & $-8.206$ & 317$\pm$9 & 267$\pm$21 & 35.96$\pm$0.13 & $+86.67\pm$0.14 & 1 & 2 \\
29 & PKSB1425-692 & 14h29m46s & $-69$d29m57.9s & 311.399 & $-8.261$ & 113$\pm$12 & 111$\pm$21 & 8.91$\pm$0.10 & $+23.64\pm$0.54 & 1 & 2 \\
30 & PKSB1442-699* & 14h47m06s & $-70$d08m4.2s & 312.521 & $-9.447$ & 496$\pm$12 & 433$\pm$11 & 22.10$\pm$0.16 & $-39.34\pm$0.21 & 3 & 3 \\
31 & PKSB1517-682 & 15h22m16s & $-68$d28m11.4s & 316.089 & $-9.542$ & 283$\pm$8 & 224$\pm$10 & 24.21$\pm$0.17 & $+210.84\pm$0.16 & 1 & 1 \\
32 & SPASSJ160700-452802 & 16h07m00s & $-45$d28m3.0s & 334.962 & $4.894$ & 806$\pm$14 & 692$\pm$10 & 36.59$\pm$0.33 & $-94.27\pm$0.14 & 3 & 3 \\
33 & SPASSJ161717-584807 & 16h17m17s & $-58$d48m7.0s & 326.969 & $-5.901$ & 3777$\pm$54 & 4276$\pm$108 & 36.53$\pm$0.60 & $+74.09\pm$0.23 & 2 & 2 \\
34 & NVSSJ165956-305205 & 16h59m55s & $-30$d52m7.0s & 352.801 & $7.078$ & 540$\pm$16 & 434$\pm$16 & 1.22$\pm$0.16 & $+151.50\pm$3.85 & 2 & 3 \\
35 & SPASSJ134355-564917 & 13h43m56s & $-56$d49m17.1s & 310.101 & $5.318$ & 362$\pm$12 & 309$\pm$28 & 33.31$\pm$0.16 & $+113.30\pm$0.15 & 1 & 3 \\
36 & SPASSJ135734-532128 & 13h57m34s & $-53$d21m28.3s & 312.813 & $8.24$ & 392$\pm$8 & 356$\pm$9 & 50.41$\pm$0.17 & $-6.82\pm$0.08 & 3 & 3 \\
37 & SPASSJ163305-450926* & 16h33m05s & $-45$d09m26.9s & 338.442 & $1.887$ & 514$\pm$23 & 392$\pm$30 & 25.52$\pm$0.31 & $-209.79\pm$0.20 & 3 & 4 \\
38 & NVSSJ161809-391811 & 16h18m08s & $-39$d18m14.0s & 340.733 & $7.946$ & 570$\pm$14 & 517$\pm$16 & 15.56$\pm$0.23 & $+128.50\pm$0.47 & 2 & 2 \\
39 & PKSB1508-649 & 15h12m34s & $-65$d06m47.8s & 317.085 & $-6.195$ & 467$\pm$7 & 440$\pm$6 & 8.93$\pm$0.23 & $-103.82\pm$0.50 & 1 & 3 \\
40 & SPASSJ144106-561659 & 14h41m06s & $-56$d16m59.6s & 317.768 & $3.397$ & 886$\pm$31 & 709$\pm$60 & 48.20$\pm$0.17 & $-79.36\pm$0.12 & 1 & 2 \\
41 & SPASSJ144233-540746 & 14h42m33s & $-54$d07m46.2s & 318.85 & $5.272$ & 432$\pm$11 & 412$\pm$12 & 19.37$\pm$0.12 & $+47.64\pm$0.21 & 2 & 2 \\
42 & NVSSJ172242-281953 & 17h22m42s & $-28$d19m57.0s & 357.78 & $4.554$ & 870$\pm$12 & 794$\pm$12 & 26.90$\pm$0.42 & $+97.56\pm$0.21 & 3 & 3 \\
43 & SPASSJ163158-593500 & 16h31m58s & $-59$d35m0.2s & 327.721 & $-7.796$ & 481$\pm$34 & ... & 9.27$\pm$0.12 & $+116.47\pm$0.44 & 2 & 3 \\
44 & SPASSJ145506-494805* & 14h55m07s & $-49$d48m5.8s & 322.515 & $8.309$ & 612$\pm$14 & 550$\pm$20 & 15.27$\pm$0.10 & $-9.93\pm$0.24 & 2 & 4 \\
45 & NVSSJ172337-272843 & 17h23m35s & $-27$d28m37.0s & 358.603 & $4.869$ & 52$\pm$3 & 42$\pm$6 & 2.61$\pm$0.14 & $-54.74\pm$1.69 & 2 & 2 \\
46 & SPASSJ151108-520350 & 15h11m08s & $-52$d03m50.2s & 323.626 & $5.104$ & 222$\pm$3 & 209$\pm$6 & 7.73$\pm$0.15 & $-130.34\pm$0.41 & 2 & 2 \\
47 & SPASSJ151739-510634 & 15h17m39s & $-51$d06m34.4s & 324.994 & $5.389$ & 139$\pm$6 & 151$\pm$17 & 5.29$\pm$0.15 & $-82.71\pm$0.83 & 1 & 1 \\
48 & NVSSJ172836-271236 & 17h28m34s & $-27$d12m41.0s & 359.441 & $4.097$ & 437$\pm$6 & 373$\pm$12 & 5.11$\pm$0.30 & $-60.90\pm$1.63 & 5 & 5 \\
49 & SPASSJ152154-531348 & 15h21m55s & $-53$d13m49.0s & 324.4 & $3.251$ & 280$\pm$12 & 222$\pm$26 & 6.45$\pm$0.11 & $-20.22\pm$0.70 & 1 & 3 \\
50 & SPASSJ152426-483956 & 15h24m26s & $-48$d39m56.3s & 327.252 & $6.846$ & 404$\pm$41 & 347$\pm$63 & 10.45$\pm$0.37 & $-199.65\pm$1.21 & 2 & 3 \\
51 & SPASSJ152436-643159* & 15h24m36s & $-64$d31m59.6s & 318.485 & $-6.392$ & 493$\pm$13 & 430$\pm$14 & 20.89$\pm$0.21 & $+183.48\pm$0.21 & 1 & 2 \\
52 & SPASSJ153023-605032 & 15h30m23s & $-60$d50m32.3s & 321.12 & $-3.714$ & 147$\pm$4 & 126$\pm$8 & 10.60$\pm$0.14 & $-144.69\pm$0.36 & 1 & 4 \\
53 & SPASSJ153419-535113 & 15h34m19s & $-53$d51m13.2s & 325.575 & $1.698$ & 518$\pm$4 & 491$\pm$8 & 11.62$\pm$0.18 & $-52.13\pm$0.31 & 1 & 4 \\
54 & PKSB1653-331 & 16h56m39s & $-33$d11m37.0s & 350.526 & $6.196$ & 420$\pm$22 & 396$\pm$26 & 40.28$\pm$0.17 & $-82.56\pm$0.12 & 2 & 3 \\
55 & SPASSJ140555-565946* & 14h05m55s & $-56$d59m46.3s & 312.978 & $4.416$ & 713$\pm$16 & 598$\pm$24 & 28.98$\pm$0.23 & $-74.08\pm$0.16 & 2 & 2 \\
56 & SPASSJ154607-465519 & 15h46m07s & $-46$d55m19.9s & 331.247 & $6.12$ & 531$\pm$9 & 482$\pm$7 & 18.31$\pm$0.18 & $+71.33\pm$0.22 & 1 & 2 \\
57 & SPASSJ155707-412646* & 15h57m07s & $-41$d26m46.0s & 336.293 & $9.095$ & 418$\pm$10 & 361$\pm$15 & 33.13$\pm$0.12 & $+9.76\pm$0.10 & 2 & 3 \\
58 & SPASSJ143353-533736 & 14h33m53s & $-53$d37m36.2s & 317.878 & $6.246$ & 221$\pm$8 & 189$\pm$12 & 2.83$\pm$0.15 & $+77.68\pm$1.24 & 1 & 2 \\
59 & NVSSJ162743-395303 & 16h27m43s & $-39$d53m7.0s & 341.615 & $6.219$ & 1305$\pm$24 & 1128$\pm$10 & 9.31$\pm$0.30 & $-50.04\pm$0.74 & 3 & 3 \\
60 & SPASSJ145739-650240 & 14h57m39s & $-65$d02m40.7s & 315.745 & $-5.362$ & 259$\pm$8 & 215$\pm$11 & 12.18$\pm$0.12 & $-42.07\pm$0.41 & 1 & 2 \\
61 & SPASSJ161255-573638 & 16h12m55s & $-57$d36m38.9s & 327.384 & $-4.645$ & 830$\pm$48 & 360$\pm$32 & 17.10$\pm$0.14 & $-18.32\pm$0.24 & 1 & 4 \\
62 & SPASSJ161432-562236* & 16h14m32s & $-56$d22m36.6s & 328.397 & $-3.903$ & 1529$\pm$28 & 1401$\pm$39 & 42.43$\pm$0.31 & $-29.45\pm$0.10 & 3 & 3 \\
63 & SPASSJ154523-452432 & 15h45m23s & $-45$d24m32.9s & 332.088 & $7.388$ & 170$\pm$14 & 143$\pm$15 & 11.67$\pm$0.13 & $-64.49\pm$0.35 & 1 & 2 \\
64 & SPASSJ161451-434125 & 16h14m50s & $-43$d41m25.0s & 337.204 & $5.238$ & 512$\pm$16 & 474$\pm$35 & 4.63$\pm$0.11 & $-0.04\pm$0.92 & 1 & 4 \\
65 & SPASSJ161551-633314 & 16h15m51s & $-63$d33m14.4s & 323.488 & $-9.172$ & 693$\pm$29 & 532$\pm$72 & 8.04$\pm$0.25 & $+200.80\pm$0.61 & 2 & 4 \\
66 & SPASSJ162941-422659 & 16h29m41s & $-42$d26m59.0s & 340.001 & $4.187$ & 205$\pm$4 & 181$\pm$5 & ... & $+437.73\pm$5.35 & 1 & 4 \\
67 & SPASSJ161950-455109 & 16h19m50s & $-45$d51m9.7s & 336.325 & $3.072$ & 646$\pm$19 & 521$\pm$23 & 71.70$\pm$0.28 & $-81.23\pm$0.08 & 2 & 4 \\
68 & SPASSJ162202-430822 & 16h22m02s & $-43$d08m22.0s & 338.525 & $4.716$ & 984$\pm$26 & 802$\pm$54 & 24.97$\pm$0.27 & $+80.83\pm$0.22 & 3 & 3 \\
69 & SPASSJ162346-564509 & 16h23m46s & $-56$d45m9.8s & 329.04 & $-5.065$ & 582$\pm$13 & 485$\pm$6 & 10.34$\pm$0.24 & $+69.46\pm$0.64 & 2 & 2 \\
70 & SPASSJ161441-442447* & 16h14m41s & $-44$d24m47.5s & 336.682 & $4.736$ & 1005$\pm$15 & 918$\pm$12 & 52.20$\pm$0.25 & $+127.38\pm$0.08 & 2 & 4 \\
71 & SPASSJ162628-580020 & 16h26m28s & $-58$d00m20.3s & 328.389 & $-6.197$ & 330$\pm$16 & 252$\pm$43 & 11.16$\pm$0.15 & $+49.93\pm$0.36 & 1 & 2 \\
72 & SPASSJ162650-421129 & 16h26m50s & $-42$d11m29.0s & 339.824 & $4.746$ & 714$\pm$100 & ... & 15.24$\pm$0.28 & $+6.32\pm$0.52 & 2 & 3 \\
73 & SPASSJ162701-604320 & 16h27m01s & $-60$d43m20.1s & 326.456 & $-8.122$ & ... & 78$\pm$15 & 3.89$\pm$0.12 & $+98.83\pm$0.93 & 1 & 2 \\
74 & NVSSJ173620-283552 & 17h36m18s & $-28$d36m2.0s & 359.205 & $1.907$ & 183$\pm$4 & 156$\pm$7 & 14.76$\pm$0.21 & $-348.57\pm$0.35 & 4 & 5 \\
75 & SPASSJ163056-580102* & 16h30m56s & $-58$d01m2.3s & 328.789 & $-6.635$ & 292$\pm$8 & 246$\pm$7 & 5.70$\pm$0.08 & $+89.20\pm$0.66 & 1 & 3 \\
76 & SPASSJ163958-603651 & 16h39m58s & $-60$d36m51.8s & 327.612 & $-9.227$ & 638$\pm$56 & 207$\pm$22 & 10.94$\pm$0.15 & $+97.14\pm$0.43 & 2 & 3 \\
\hline
    \end{tabular}
    \end{adjustbox}
    \label{tab:cat}
\end{table*}

\begin{table*}
    \setcounter{table}{0}
    \renewcommand{\thetable}{\ref{tab:cat} Continued}
    \centering
    \caption{}
    \begin{adjustbox}{width=\textwidth,center}
    \begin{tabular}{llllllllllcc}
    \hline
    & \citetalias{Schnitzeler_2019} ID (1) & RA (2) & Dec (3) & $l$ (4) & $b$ (5) & $I_\mathrm{MFS}$ (6) & $I_{2100}$ (7) & PI (8)& $\phi_\mathrm{peak}$ (9) & $n$ (10) & $n$ (11)\\
    & & (J2000) & (J2000) & (deg) & (deg) & (mJy) & (mJy) & (mJy) & rad m$^{-2}$ & (this work) & (\citetalias{Schnitzeler_2019}) \\
    \hline
77 & SPASSJ165908-532028* & 16h59m08s & $-53$d20m28.0s & 334.961 & $-6.644$ & 538$\pm$12 & 465$\pm$11 & 38.41$\pm$0.17 & $+118.12\pm$0.12 & 1 & 2 \\
78 & PKSB1637-319* & 16h40m38s & $-32$d04m37.0s & 349.235 & $9.51$ & 327$\pm$5 & 297$\pm$6 & 27.81$\pm$0.12 & $-33.44\pm$0.14 & 1 & 1 \\
79 & SPASSJ151439-474835 & 15h14m40s & $-47$d48m30.0s & 326.338 & $8.448$ & 1082$\pm$11 & 1216$\pm$10 & 10.46$\pm$0.16 & $-7.68\pm$0.30 & 1 & 1 \\
80 & SPASSJ171405-515258 & 17h14m05s & $-51$d52m58.0s & 337.501 & $-7.605$ & 84$\pm$34 & 294$\pm$24 & 16.69$\pm$0.16 & $-50.08\pm$0.27 & 1 & 3 \\ 
81 & SPASSJ171510-515708 & 17h15m09s & $-51$d57m8.0s & 337.541 & $-7.781$ & 440$\pm$18 & 270$\pm$33 & 11.11$\pm$0.19 & $-50.43\pm$0.42 & 1 & 3 \\
82 & SPASSJ172135-514628 & 17h21m34s & $-51$d46m28.0s & 338.253 & $-8.499$ & 384$\pm$18 & 298$\pm$30 & 7.92$\pm$0.15 & $+43.84\pm$0.54 & 1 & 3 \\
83 & SPASSJ172139-495033 & 17h21m38s & $-49$d50m33.0s & 339.879 & $-7.432$ & 130$\pm$30 & 144$\pm$39 & 3.19$\pm$0.15 & $-70.85\pm$1.47 & 1 & 3 \\
84 & SPASSJ172315-463135 & 17h23m14s & $-46$d31m35.0s & 342.798 & $-5.8$ & 516$\pm$40 & 266$\pm$36 & 38.81$\pm$0.16 & $-91.50\pm$0.13 & 1 & 3 \\
85 & SPASSJ172456-444054 & 17h24m56s & $-44$d40m54.0s & 344.5 & $-5.015$ & 284$\pm$6 & 249$\pm$7 & 13.55$\pm$0.14 & $+161.11\pm$0.32 & 1 & 1 \\
86 & SPASSJ173219-482740 & 17h32m19s & $-48$d27m40.0s & 342.008 & $-8.139$ & 415$\pm$15 & 231$\pm$19 & 8.79$\pm$0.14 & $-235.78\pm$0.52 & 2 & 4 \\
87 & SPASSJ173720-412636 & 17h37m19s & $-41$d26m36.0s & 348.465 & $-5.154$ & 317$\pm$31 & 377$\pm$83 & 10.66$\pm$0.13 & $-108.20\pm$0.52 & 1 & 2 \\
88 & SPASSJ174304-405729 & 17h43m03s & $-40$d57m29.0s & 349.448 & $-5.817$ & 408$\pm$6 & 395$\pm$9 & 19.04$\pm$0.13 & $-24.43\pm$0.23 & 1 & 3 \\
89 & SPASSJ175548-443006* & 17h55m47s & $-44$d30m6.0s & 347.534 & $-9.615$ & 407$\pm$11 & 363$\pm$11 & 7.88$\pm$0.18 & $+49.90\pm$0.74 & 1 & 2 \\
90 & SPASSJ143012-545251 & 14h30m12s & $-54$d52m51.9s & 316.898 & $5.29$ & 373$\pm$21 & 259$\pm$32 & 5.04$\pm$0.13 & $-176.49\pm$0.96 & 1 & 2 \\
91 & NVSSJ180242-394005 & 18h02m42s & $-39$d40m8.0s & 352.451 & $-8.425$ & 1270$\pm$4 & 1250$\pm$5 & 14.62$\pm$0.21 & $+90.88\pm$0.18 & 1 & 1 \\
92 & PKSB1654-339* & 16h57m07s & $-34$d00m0.0s & 349.947 & $5.623$ & 112$\pm$3 & 117$\pm$8 & 7.78$\pm$0.14 & $-142.62\pm$0.53 & 1 & 4 \\
93 & PKSB1657-261 & 17h00m57s & $-26$d10m28.0s & 356.716 & $9.74$ & 1009$\pm$8 & 1059$\pm$12 & 25.13$\pm$0.21 & $-27.97\pm$0.22 & 2 & 2 \\
94 & PKSB1657-340 & 17h00m15s & $-34$d09m22.0s & 350.227 & $5.013$ & 585$\pm$9 & 498$\pm$9 & 6.91$\pm$0.20 & $-132.48\pm$0.68 & 4 & 4 \\
95 & SPASSJ170758-520057 & 17h07m58s & $-52$d00m57.0s & 336.838 & $-6.916$ & 96$\pm$3 & ... & 8.01$\pm$0.14 & $-106.38\pm$0.53 & 1 & 3 \\ 
\hline
    \end{tabular}
    \end{adjustbox}
    \label{tab:cat2}
\end{table*}

\end{document}